\documentclass[12pt]{article}
\usepackage{scicite}
\usepackage{times}
\usepackage{graphicx}
\usepackage{comment}
\usepackage{ifthen}
\usepackage{amsmath}
\usepackage{amsfonts}
\usepackage{amssymb}
\usepackage{subfigure}
\usepackage{epsfig}
\newcommand{\luna}{{\tt LUNA}}
\newcommand{\multi}{{\sc MultiNest}}

\newcommand{\koi}{KOI-872}
\newcommand{\koione}{KOI-872.01}
\newcommand{\koitwo}{KOI-872.02}
\newcommand{\koithree}{KOI-872.03}
\newcommand{\hcv}{HCV/HCA-439}
\newcommand{\hcvb}{HCV/HCA-439.01}

\newcommand{\koib}{KOI-872b}
\newcommand{\koic}{KOI-872c}
\newcommand{\farcs}{\mbox{\ensuremath{.\!\!^{\prime\prime}}}}

\newenvironment{sciabstract}{%
\begin{quote} \bf}
{\end{quote}}

\renewcommand\refname{References and Notes}

\newcounter{lastnote}

\title{The Detection and Characterization of a Nontransiting Planet by Transit 
Timing Variations}
\author
{David Nesvorn\'y$^{1\ast}$, David M. Kipping$^{2,3}$, Lars A. Buchhave$^{4,5}$,\\ 
 G\'asp\'ar \'A. Bakos$^{6}$, Joel Hartman$^{6}$, Allan R. Schmitt$^{7}$ \\
\normalsize{$^{1}$Department of Space Studies, Southwest Research Institute, 
Boulder, CO~80302, USA}\\
\normalsize{$^{2}$Harvard-Smithsonian Center for Astrophysics, Cambridge, 
MA~02138, USA}\\
\normalsize{$^{3}$NASA Carl Sagan Fellow}\\
\normalsize{$^{4}$Niels Bohr Institute, University of Copenhagen, DK-2100, Copenhagen, Denmark}\\
\normalsize{$^{5}$Centre for Star and Planet Formation, Natural History Museum of Denmark}\\
\normalsize{University of Copenhagen, DK-1350, Copenhagen, Denmark}\\
\normalsize{$^{6}$Department of Astrophysical Sciences, Princeton University, 
Princeton, NJ~05844, USA}\\
\normalsize{$^{7}$Citizen Science}\\
\\
\normalsize{$^{\ast\ast}$ To whom correspondence should be addressed; E-mail: 
davidn@boulder.swri.edu.}
}

\date{}

\begin{document} 

\baselineskip23pt

\maketitle 

\begin{sciabstract}

The Kepler Mission is monitoring the brightness of $\sim$150,000 stars 
searching for evidence of planetary transits. As part of the ``Hunt for Exomoons 
with Kepler'' (HEK) project, we report a planetary system with two confirmed 
planets and one candidate planet discovered using the publicly available data 
for \koi. Planet b transits the host star with a period $P_b=33.6$\,d and 
exhibits large transit timing variations indicative of a perturber. Dynamical 
modeling uniquely detects an outer nontransiting planet c near the 5:3 resonance 
($P_c=57.0$\,d) of mass 0.37 times that of Jupiter. Transits of a third 
planetary candidate are also found: a 1.7-Earth radius super-Earth with a 6.8\,d 
period. Our analysis indicates a system with nearly coplanar and circular 
orbits, reminiscent of the orderly arrangement within the solar system.

\end{sciabstract}

If a planet's orbit is viewed nearly edge-on, the planet may transit over the 
disk of its host star and periodically block a small fraction of the starlight. 
The planet's presence is then revealed by a small and repetitive decrease of the 
host star's brightness during transits. The transit light curve is characterized 
by the time of transit minimum $\tau$, the transit depth $\delta$, the total 
duration $T_{14}$, and the partial duration $T_{23}$ \cite{carter:2008}. A 
precise measurement of these terms allows an observer to infer the 
physical properties of the system, such as the radius ratio, $p=R_P/R_*$, 
transit impact parameter, $b_P$, and scaled semi-major axis, $a_P/R_*$, where 
$R_*$ is the star's physical radius and the subscript $P$ denotes ``planet''.

For a planet following a strictly Keplerian orbit, the spacing, timing and other 
properties of the transit light curve should be unchanging in time. Several 
effects, however, can produce deviations from the Keplerian case so that the 
spacing of $\tau$ is not strictly periodic and/or $T_{14}$ varies from 
transit to transit. Such changes are known as transit timing variations (TTVs) 
and transit duration variations (TDVs), respectively. TTVs are particularly 
sensitive to gravitational perturbations from additional planets orbiting the 
host star \cite{miralda:2002,agol:2005,holman:2005} and distant large moons 
orbiting the transiting planet \cite{sart:1999,kipping:2009a,kipping:2009b}.

As part of the ``Hunt for Exomoons with Kepler'' (HEK) project \cite{hek:2012}, 
we analyzed the publicly available Kepler data up to Quarter 6 (Q6; released on 
January~7, 2012). At the time of writing, the 33.6-day-period planetary 
candidate Kepler-object-of-interest 872.01 (\koione) is the only known candidate 
in the system \cite{borucki:2011}. The candidate was identified through HEK's 
target selection procedure as a high priority object because of the presence of 
visual transit anomalies and the dynamical capacity to host a moon.

\begin{figure}
\centering
\includegraphics[width=18.0 cm]{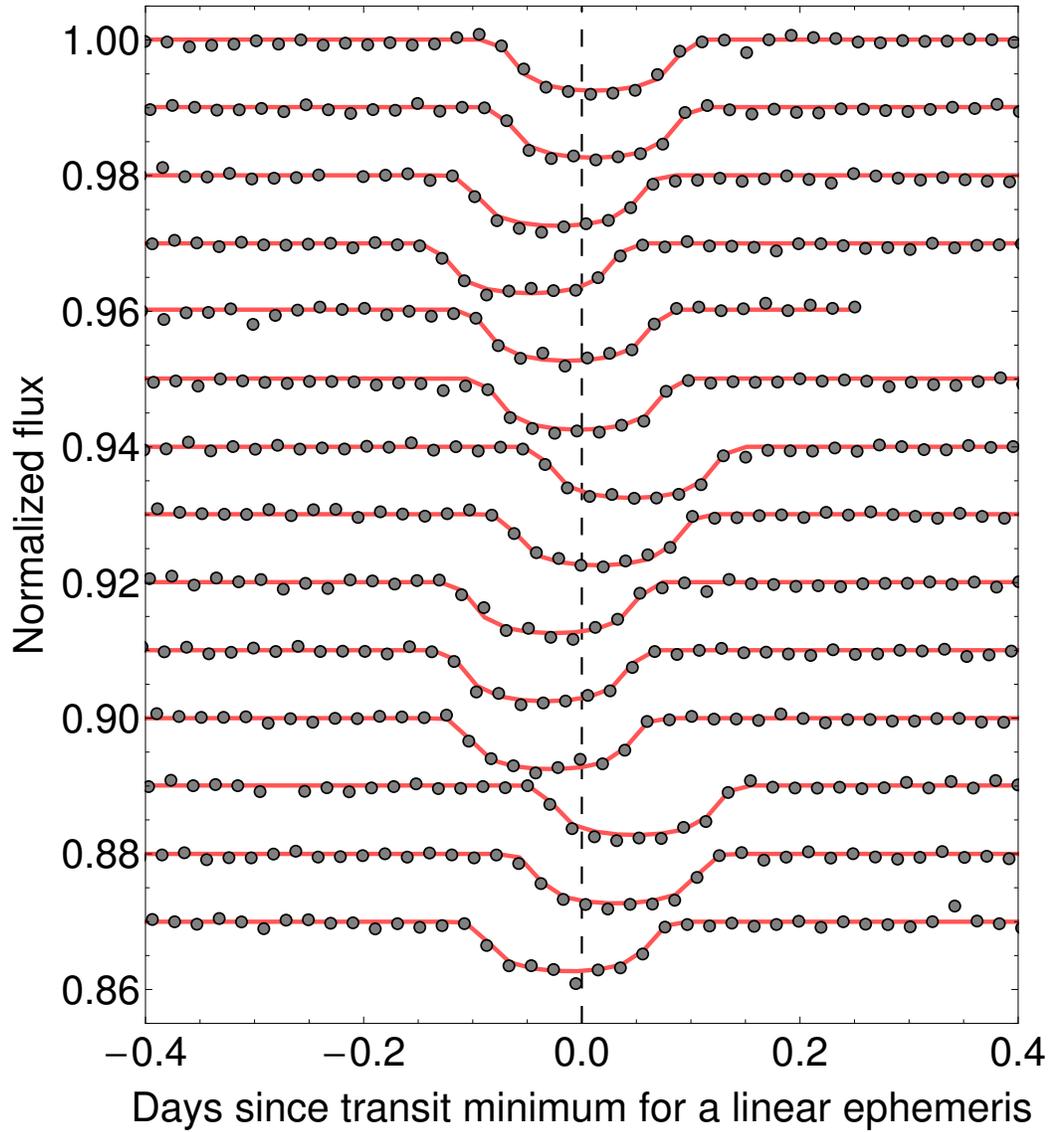}
\caption{Maximum likelihood transit model (red line) overlaid with the 
long-cadence Kepler offsetted data for \koib. The large TTVs are evident 
visually from the light curve. The ramp-affected transit is excluded here
(see Fig.~S2).}
\label{fig:lightcurves}
\end{figure}

We detrended the raw Kepler photometry of \koi, covering 15 transits, using a 
harmonic filter and an exponential decay ramp correction \cite{som}. We tested 
several models to explain the photometry using the multimodal nested sampling 
algorithm \multi\ \cite{feroz:2008,feroz:2009}, designed to compute the Bayesian 
evidence for each model. The favored model was found to be that of a planet 
undergoing TTVs (each transit has a unique $\tau$ but common parameters for all 
other terms; model $\mathcal{M}_T$). This model is preferred over that of a 
planet on a linear ephemeris, $\mathcal{M}_P$, at a confidence of 
$43.7$\,$\sigma$ (Fig.~\ref{fig:lightcurves}).

\begin{figure}
\centering
\includegraphics[width=14.cm]{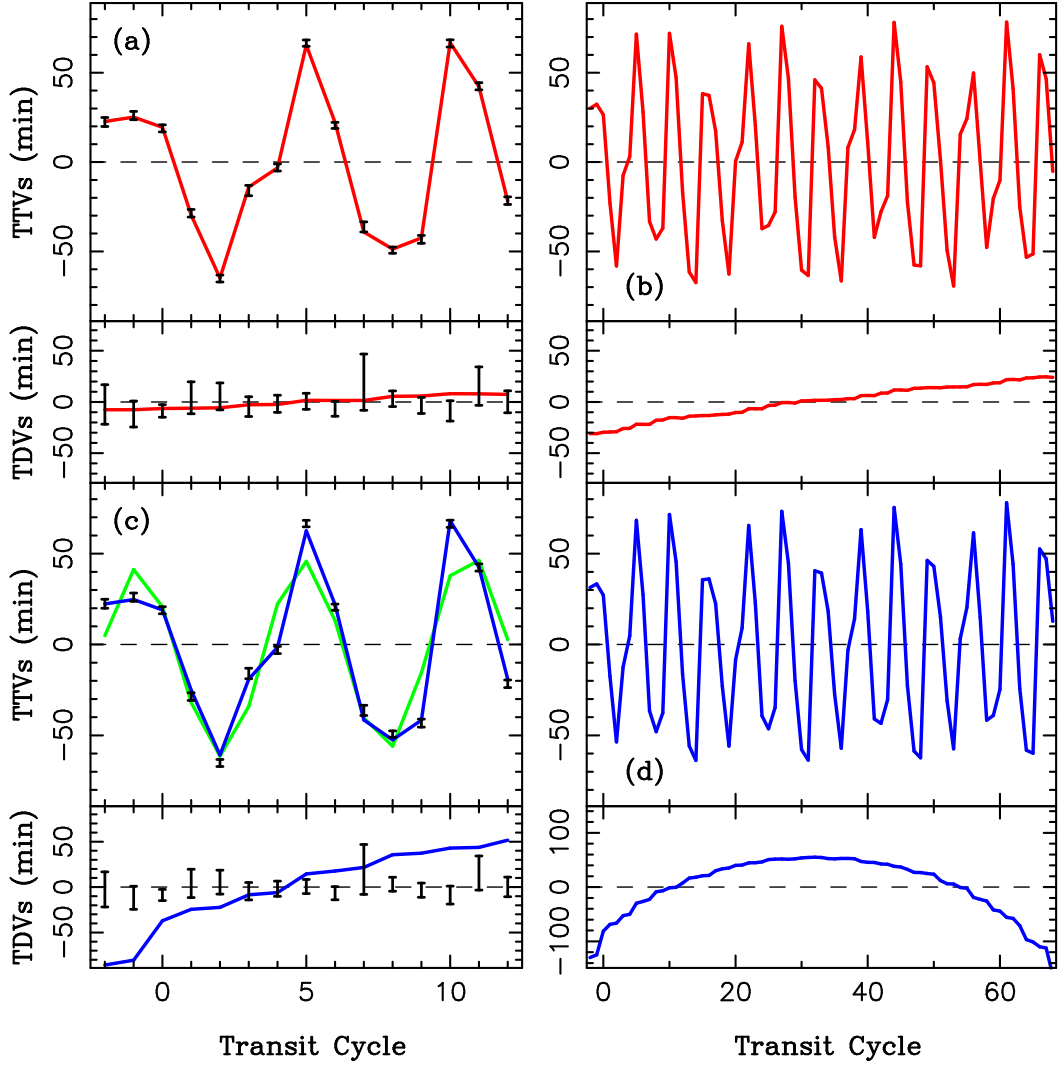}
\caption{Transit timings variations. The measured TTVs (from model 
$\mathcal{M}_T$) and TDVs (from model $\mathcal{M}_V$) and their 
uncertainties are indicated in panels (a) and (c). Panel (a) shows the 
calculated values for s1 (red line). The TDVs of s1 are consistent with the 
measured, flat TDV profile. Panel (c) shows s2 (blue line) and our best moon model
(green line). While s2 also fits the measured TTVs relatively well, the 
strong TDV trend in (c) is inconsistent with the measurements. Panels (b) and 
(d) show our predictions for s1 and s2.}
\label{fig:TTVs}
\end{figure}

We computed TTVs relative to the linear ephemeris derived from $\mathcal{M}_P$ 
($P_b = 33.60134 \pm 0.00021$\,d and $\tau_0 = 2455053.2826\pm0.0014$\,
BJD$_{\mathrm{UTC}}$, where BJD$_{\mathrm{UTC}}$ is understood to be barycentric
Julian date in coordinated universal time). The results (Fig.~\ref{fig:TTVs}) 
indicate that \koione\ exhibits large (2\,hours) and complex TTVs with a 
dominant period of about $\simeq$5.6 transit cycles ($\simeq$190\,d). These are 
among the largest TTVs ever detected. A model including TDVs is disfavored 
relative to the TTV-only model at a confidence of $17.5$\,$\sigma$.

Because a moon should induce TDVs and TTVs, the lack of TDVs does not favor a 
moon hypothesis. Further, Q4-Q6 photometry show complex stellar activity, which 
may be responsible for the initially identified visual anomalies. In support of 
these arguments, we found a planet-with-moon model ($\mathcal{M}_M$) inadequate 
to explain the measured TTVs (Fig.~\ref{fig:TTVs}c). 

A distant stellar companion or secular/resonant perturbations from a planetary 
companion also cannot explain \koione's TTVs, because these effects would 
produce sinusoidal patterns with a very long period \cite{montalto:2010}. 
Moreover, other known TTV effects, such as parallax effects \cite{scharf:2007}, 
the Applegate effect \cite{watson:2010}, and stellar proper motion 
\cite{rafikov:2009}, can be ruled out because they are unable to produce 
such a large TTV amplitude. 

We applied the TTV inversion method described in \cite{nesvorny:2010} to test 
whether the observed TTVs are consistent with short-period perturbations from a 
planetary companion and whether a unique set of parameters can be derived to 
describe the physical and orbital properties of that companion. The short-period 
perturbations are small variations around the mean Keplerian orbit of a planet 
with characteristic periods comparable to the planet's orbital period. The 
inversion method is based on perturbation theory, which greatly speeds up the 
computation of the timing and duration of individual transits 
\cite{nesvorny:2008}.

We tested orbits with periods between 1\,day and 10\,years, including the cases of 
highly eccentric and/or retrograde planets \cite{trojan}. The identified solutions were fine 
tuned using a precise $N$-body integrator. We used the downhill simplex method 
to search for the minimum of $\chi^2=\sum_{j=-1}^{n} (\delta t_{{\rm O},j} - 
\delta t_{{\rm C},j})^2/\sigma_j^2$, where $n=15$ is the number of transits, 
$\delta t_{{\rm O},j}$ and $\delta t_{{\rm C},j}$ are the observed and 
calculated TTVs, and $\sigma_j$ is the uncertainty of $\delta t_{{\rm O},j}$.
Local minima in $\chi^2$ were tested for physical plausibility, including a 
stability test where the orbits were tracked over $10^9$\,yr with a symplectic 
integrator \cite{duncan:1998}.

All except two solutions can be ruled out because they are either dynamically 
unstable or have $\chi_{\rm min}^2>60$. The best-fit solution (hereafter s1) 
fits the data extremely well: $\chi_{\rm min,s1}^2=3.4$ for $n-m=15-7=8$ degrees of 
freedom (DOF), where $m=7$ parameters of the perturbing planet are the mass 
ratio $M_c/M_*$, period $P_c$, eccentricity $e_c$, inclination $I_c$, nodal 
longitude $\Omega_c$, pericenter longitude $\varpi_c$, and mean longitude 
$\lambda_c$. The inclination $I$ is defined relative to the reference plane that is 
90$^\circ$ tilted to the sky plane, and rotated so that $\Omega_b=270^\circ$ 
(all longitudes measured relative to the line of sight; hereafter, transit 
reference system). The orbital inclination of \koione\ relative 
to the transit plane is $I_b=0.96^\circ$, as determined from the transit fit. 

The second best-fit solution (s2) has $\chi_{\rm min,s2}^2=20.3$. It 
corresponds to a planet with $M_c \simeq 1.8\times10^{-3} M_*$, 
$P_c \simeq 81.7$\,days (i.e. $P_c/P_b = 2.43$, just inside the 5:2 resonance), 
$e_c \simeq 0.03$ and $I_c \simeq 10^\circ$. s2 can be ruled out because the 
goodness-of-fit for the TTVs is significantly worse at $\Delta \chi^2 = 16.9$; 
also, as $I_c$ is relatively large, s2 implies strong TDVs that are inconsistent 
with the observations (Fig.~\ref{fig:TTVs}c) \cite{tdv}. 

Therefore the transit variations of \koione\ can only 
be fit by s1. This was by no means expected or guaranteed because the 
short-periodic TTVs produced by the interacting planets represent only a 
very specific subset of astrophysical signals. This can be demonstrated by 
scrambling the TTV datapoints and applying the TTV inversion method to the 
scrambled data. We were unable to find a plausible planet solution for any of 
the attempted (thousand) trials with the scrambled data. This is a strong 
indication that \koi\ is a real system of at least two planets \cite{three}
(Table~\ref{tab:final}).

The scaled mass of \koitwo\ inferred from TTVs of \koione\ is 
$M_c/M_* = 3.97 \times 10^{-4}$. With $M_* \simeq 0.9\,M_{\odot}$, obtained 
from spectroscopy, this gives $M_c \simeq 0.37$\,$M_J$, or $\sim1.3$ 
Saturn masses. The mass of \koione, $M_b$, cannot be constrained from 
TTVs because the short-periodic TTVs are practically independent of $M_b$ 
\cite{agol:2005,nesvorny:2008}. The stability requirements imply that 
$M_b < 6$\,$M_J$, because the orbits are dynamically unstable with a 
more massive transiting body. This mass limit along with our vetting analysis
\cite{som} confirm \koione\ and the perturber as being genuine planets, henceforth
referred to \koib\ (corresponding to \koione) and \koic.

The period ratio, $P_c/P_b = 1.697$ indicates that the two planets are just 
outside the 5:3 resonance. This may be a 
relatively common configuration probably related to the radial migration of 
planets in the protoplanetary nebula \cite{lissauer:2011a}. The resonant angle, 
$5\lambda_c-3\lambda_b-2\varpi_c$, circulates in the retrograde sense with the 
period of $\simeq2$ yr. The dominant TTV period ($\simeq5.6$ transit cycles; 
Fig.~\ref{fig:TTVs}) comes from the relatively distant 2:1 and 3:2 terms 
(periods $\simeq190$\,yr). 

The orbital eccentricities are $e_b<0.02$ and $e_c\simeq0.015$. The nearly circular 
orbits probably indicate that the two planets formed at, or migrated to, 
their present orbits without suffering any strong dynamical instability. 
The two planets also exhibit nearly but not exactly coplanar orbits. The orbital 
inclination of \koic\ with respect to the transit plane is $I_c<4.5^\circ$ (99\% 
confidence interval) with the best TTV fit indicating $I_c=2.6^\circ$. The 
best-fit inclination value and $\Omega_c\simeq 298^\circ$ obtained from the TTVs 
suggest the planet just avoids transiting. A search for the transits of \koic\
indeed yielded no events, implying that $I_c>1^\circ$.

Although no transits of \koic\ were detected, we do detect a 9-$\sigma$ transit
signal on a short period of 6.8\,d. The transit corresponds to a
1.7\,$R_{\oplus}$ Super-Earth, which we refer to as \koithree. Because of the 
probable low-mass nature of this body, the expected TTVs induced on \koib\ would be
around 1\,s in amplitude, too small to detect with our data. Further, the
TTVs of \koithree\ itself are estimated to be $\sim10$\,s, also undetectable.
Without TTVs, we are unable to confirm that \koithree\ orbits the same star
as \koib\ and \koic, as opposed to a blended background star. However, our
analysis finds that \koithree's light curve derived stellar density is 
consistent with that of \koib\ for both planets on near-circular orbits.

We predict that the radial velocity (RV) measurements of \koi\ should reveal at least two
basic periods. The RV term with a 57.0\,d period, corresponding to \koic, should have 
$K_c\simeq20$~m~s$^{-1}$ half-amplitude. The amplitude of the 33.6\,d period term, 
corresponding to \koib, is uncertain because we do not have a good constraint on $M_b$. 
The 6.8\,d term, corresponding to \koithree, will be difficult to detect because 
$K_3\sim1$-2~m~s$^{-1}$ (assuming Earth-like density).

\begin{figure}
\centering
\includegraphics[width=14.cm]{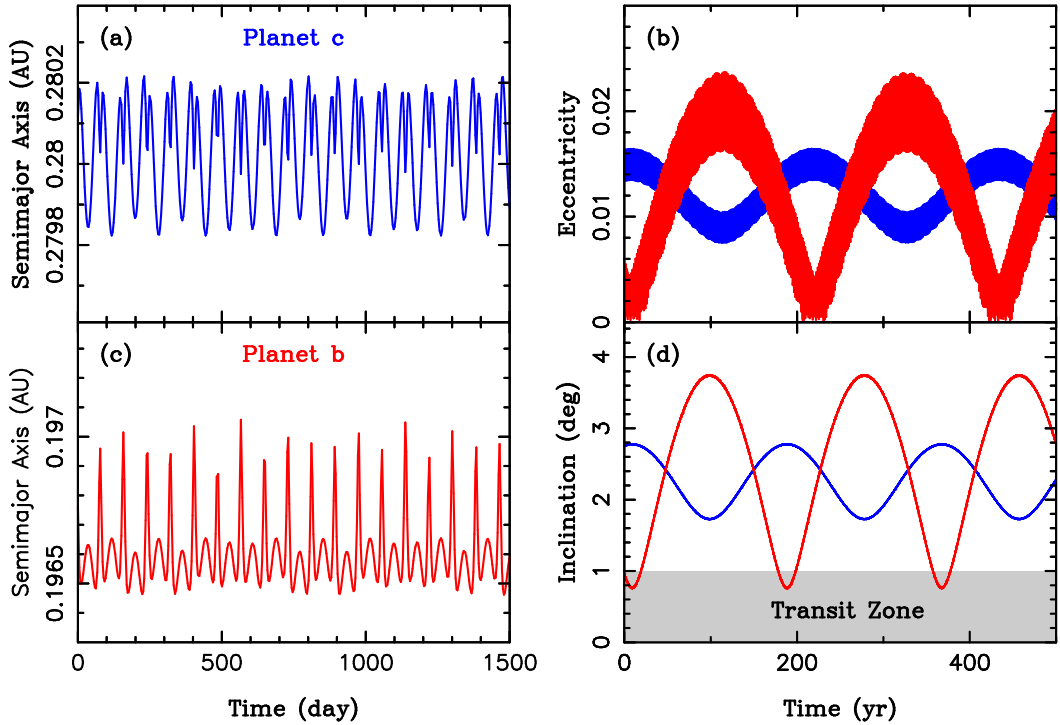}
\caption{Orbit evolution. Figure shows the semi-major axis (panels a and c), 
eccentricity (b) and inclination (d) for the best fit solution. The orbital 
elements of \koib\ and \koic\ are shown in red and blue, respectively.
The inclination in (d) is defined relative to the transit plane ($I=0^\circ$ 
would correspond to a transit across the middle of the host star's disk). An 
approximate transit zone for $\Omega=270^\circ$ is shaded. $\Omega_b$ and 
$\Omega_c$ are locked near $270^\circ$, because the invariant plane of planets 
is tilted to the transit plane. \koithree\ is omitted here because it has a 
negligible effect on the dynamical evolution of \koib\ and \koic. }
\label{fig:evolution}
\end{figure}

To gain insights into the system's dynamical behavior we numerically integrated 
the planetary orbits starting from the best-fit solution 
(Fig.~\ref{fig:evolution}).
The semi-major axis of \koib\ experiences short-period variations, related to 
the TTVs, with an amplitude of $\simeq 5\times 10^{-4}$\,AU. The 
eccentricities undergo anti-correlated oscillations, as dictated by the angular 
momentum conservation, around means $\bar{e}_b=0.013$ and 
$\bar{e}_c=0.012$, with a period of $\simeq$200\,yr (corresponding to 
$\varpi_b$'s precession period; $\varpi_c$ precesses much slower). 
The anti-correlated oscillations of inclinations relative to the transit plane 
are a geometrical effect resulting from the shared precession of $\Omega_b$ and 
$\Omega_c$ ($\simeq 200$\,yr period). The inclinations relative to 
the invariant plane are nearly constant with the mutual inclination 
$\simeq1^\circ$. If $I_b$ relative to the transit plane increases in the next 
decades (Fig.~3d), \koib's transits will gradually disappear. 

TTVs were originally proposed as a nontransiting planet detection method 
\cite{miralda:2002,agol:2005,holman:2005}, but have recently found more use in 
validating the transiting planet candidates from Kepler 
\cite{holman:2010,lissauer:2011}. Kepler has previously inferred the
presence of a nontransiting planet via TTVs, but showed that the 
measurements were unable to support a unique solution \cite{ballard:2011}. Here 
we have demonstrated the full potential of TTVs as a method to detect nontransiting planets 
and precisely characterize their properties. 


\clearpage
\begin{table}
\centering
\resizebox{!}{7cm} {

\begin{tabular}{llll}
\hline\hline 
 & \emph{\koib} & \emph{\koic} & \emph{\koithree} \\ [0.5ex]
\hline
$\tau_0$\,[BJD$_{\mathrm{UTC}}$] & $2455053.2826_{-0.0014}^{+0.0013}$ & - & $2455255.2603_{-0.0031}^{+0.0032}$ \\
$P_P$\,[days] & $33.60134_{-0.00020}^{+0.00021}$ & $57.011_{-0.061}^{+0.051}$ & $6.76671_{-0.00012}^{+0.00013}$ \\
$R_P/R_*$ & $0.0887_{-0.0012}^{+0.0010}$ & - & $0.01656_{-0.00082}^{+0.00079}$ \\
$b_P$ & $0.757_{-0.027}^{+0.022}$ & $2.9_{-1.8}^{+1.1}$ & $0.39_{-0.12}^{+0.19}$ \\
$a_P/R_*$ & $45.1_{-1.7}^{+2.1}$ & $64.1_{-2.5}^{+2.9}$ & $15.58_{-0.49}^{+0.51}$ \\
$i_P$\,[$^{\circ}$] & $89.038_{-0.067}^{+0.075}$ & $87.4_{-1.0}^{+1.6}$ & $88.55_{-0.69}^{+0.49}$ \\
$a_P$\,[AU] & $0.1968_{-0.0028}^{+0.0029}$ & $0.2799_{-0.0040}^{+0.0041}$ & $0.0679_{-0.0035}^{+0.0035}$ \\
$e_P$ & $0.01^{+0.01}_{-0.01}$ & $0.0146_{-0.0036}^{+0.0034}$ & $0$ (assumed) \\
$\Omega_P$\,[$^{\circ}$] & 270 & $298_{-36}^{+37}$ & - \\
$\varpi_P$\,[$^{\circ}$] & - & $330.0_{-9.2}^{+11.6}$ & - \\
$\lambda_P$\,[$^{\circ}$] & 0 & $338.2_{-1.4}^{+1.2}$ & - \\
$M_P/M_*$ & $<6.4 \times 10^{-3}$ & $3.97_{-0.11}^{+0.15}\times10^{-4}$ & - \\
$M_P$\,[$M_J$] & $<6$ & $0.376_{-0.019}^{+0.021}$ & - \\
$R_P$\,[$R_J$] & $0.808_{-0.043}^{+0.042}$ & - & $0.1510_{-0.0098}^{+0.0094}$ \\
$\rho_P$\,[kg\,m$^{-3}$] & $<14000$ & - & - \\
$T_{\mathrm{eq}}$\,[K] & $543_{-16}^{+16}$ & $455_{-14}^{+13}$ & $924_{-23}^{+24}$ \\
$M_{\mathrm{moon}}/M_P$ & $<0.021$ & - & - \\
\hline\hline
& \emph{\koi} & \\
\hline
$\rho_{*}$\,[kg\,m$^{-3}$] & $1530_{-170}^{+220}$ \\
$M_*$\,[$M_{\odot}$] & $0.902_{-0.038}^{+0.040}$ \\
$R_*$\,[$R_{\odot}$] & $0.938_{-0.039}^{+0.038}$ \\ 
$\log g_*$ & $4.447_{-0.035}^{+0.040}$ \\
$T_{\mathrm{eff}}$\,[K] & $5155 \pm 105$ \\
$L_*$\,[$L_{\odot}$] & $0.556_{-0.070}^{+0.078}$ \\
$M_V$ & $5.60_{-0.17}^{+0.17}$ \\
Age\,[Gyr] & $9.7_{-3.5}^{+3.7}$ \\
Distance\,[pc] & $855_{-65}^{+68}$ \\
$[\mathrm{M}/\mathrm{H}]$ & $0.41 \pm 0.10$ \\ [1ex]
\hline\hline 
\end{tabular}
	}
\caption{\koi\ system parameters. \koib\ parameters were computed from the 
weighted posteriors of a model accounting for TTVs, using \multi. Parameters 
fitted in the transit model are quoted as the median of the marginalized 
posteriors with $\pm$34.13\% credible intervals. Instrumental terms and times of 
transit minimum may be found in Table~S3. \koic\ parameters were computed from 
the fit to the \koib's TTVs. Parameters fitted in the TTV model are quoted as 
maximum likelihood with $\pm$34.13\% uncertainties computed using the 
$\Delta\chi^2 =1$ method described in \cite{brown:2001}, where the TTV errors 
have been rescaled such that $\chi_{\mathrm{reduced}}^2=1$. The 99\% confidence 
areas from the TTV fit alone are shown in Fig.~S10. The measured TDVs do not 
offer a meaningful constraint, since the parameter sets near s1 that fit the 
TTVs also fit the TDVs. Orbital longitudes of \koic\ are relative to the transit 
reference system on 2455053.2839\,BJD$_{\mathrm{UTC}}$. \koithree\ parameters 
were computed from an MCMC run. Moon mass constraint (3\,$\sigma$ limit) was
derived from model $\mathcal{M}_{MT2,R0}$ \cite{som}.}
\label{tab:final} 
\end{table}

\clearpage

\clearpage

\clearpage

\clearpage

\topmargin 0.0cm
\oddsidemargin 0.2cm
\textwidth 16cm 
\textheight 21cm
\footskip 1.0cm

\renewcommand\refname{References and Notes}


\setcounter{figure}{0}
\setcounter{table}{0}
\renewcommand{\thefigure}{S\arabic{figure}}
\renewcommand{\thetable}{S\arabic{table}}

\paragraph*{\Huge Supporting Online Material}

\paragraph*{System Identification}

We here discuss why \koi\ was selected for analysis by the HEK project.
The HEK project filters the list of all known \emph{Kepler} transiting planet
candidates down to a subset of the most promising for detecting exomoons. This 
process is described as ``Target Selection'' (TS) and features three distinct 
pathways: target selection visual (TSV), target selection automatic (TSA) and 
target selection opportunities (TSO). We direct the reader to 
\cite{hek:2012} for details on each procedure.

\koib, internally labeled as \hcvb, was identified independently by TSV 
and TSA as being an excellent candidate for exomoon follow-up and thus was 
prioritized for more detailed analysis. These determinations were initially 
based upon the Q1-Q3 long-cadence data only. Subsequent Q4-Q6 data is included
in the analysis presented here.

\paragraph*{Data Handling}

We will here describe the sequential steps we took in processing the 
\emph{Kepler} photometry for the target \koi.

{\it Data acquisition.}
We make use of the publicly available archival data from the 
\emph{Kepler Mission} on the MAST website, which consists of quarters 1 through
6 (Q1-Q6). All data is in long-cadence mode with nearly continuous coverage.
Full details on the data processing pipeline can be found in the data release 
handbooks and is not repeated here.

{\it Long-term detrending with a cosine filter.}
We make use of the ``raw'' (labeled as ``SAP\_FLUX'' in the header) data 
processed by the \emph{Kepler} DAWG (Data Analysis Working Group) pipeline.
All of the publicly available photometry was acquired in long-cadence (LC) mode
spanning quarters 1 through 6 (Q1-Q6) and data releases 4 through 9 (DR4-DR9).
A detailed description of the pipeline can be found in the accompanying release
notes. The ``raw'' data has been processed using PA (Photometric Analysis), 
which includes cleaning of cosmic ray hits, Argabrightenings, removal of 
background flux, aperture photometry and computation of flux-weighted centroid 
positions.

The data releases also include corrected fluxes (labeled as ``PDCSAP\_FLUX'' 
in the header), which are outputted from the PDC (Pre-search Data Conditioning) 
algorithm developed by the DAWG . As discussed in 
DR5, this data is not recommended for scientific use, owing to, in part, the 
potential for under/over-fitting of the systematic effects.
For the sake of brevity, we do not reproduce the details of the PA and PDC
steps here, but direct those interested to \cite{gilliland:2010} and the 
DR handbooks.

\begin{figure}
\subfigure[Quarter 1
\label{fig:HCV439_Q1raw}]
{\epsfig{file=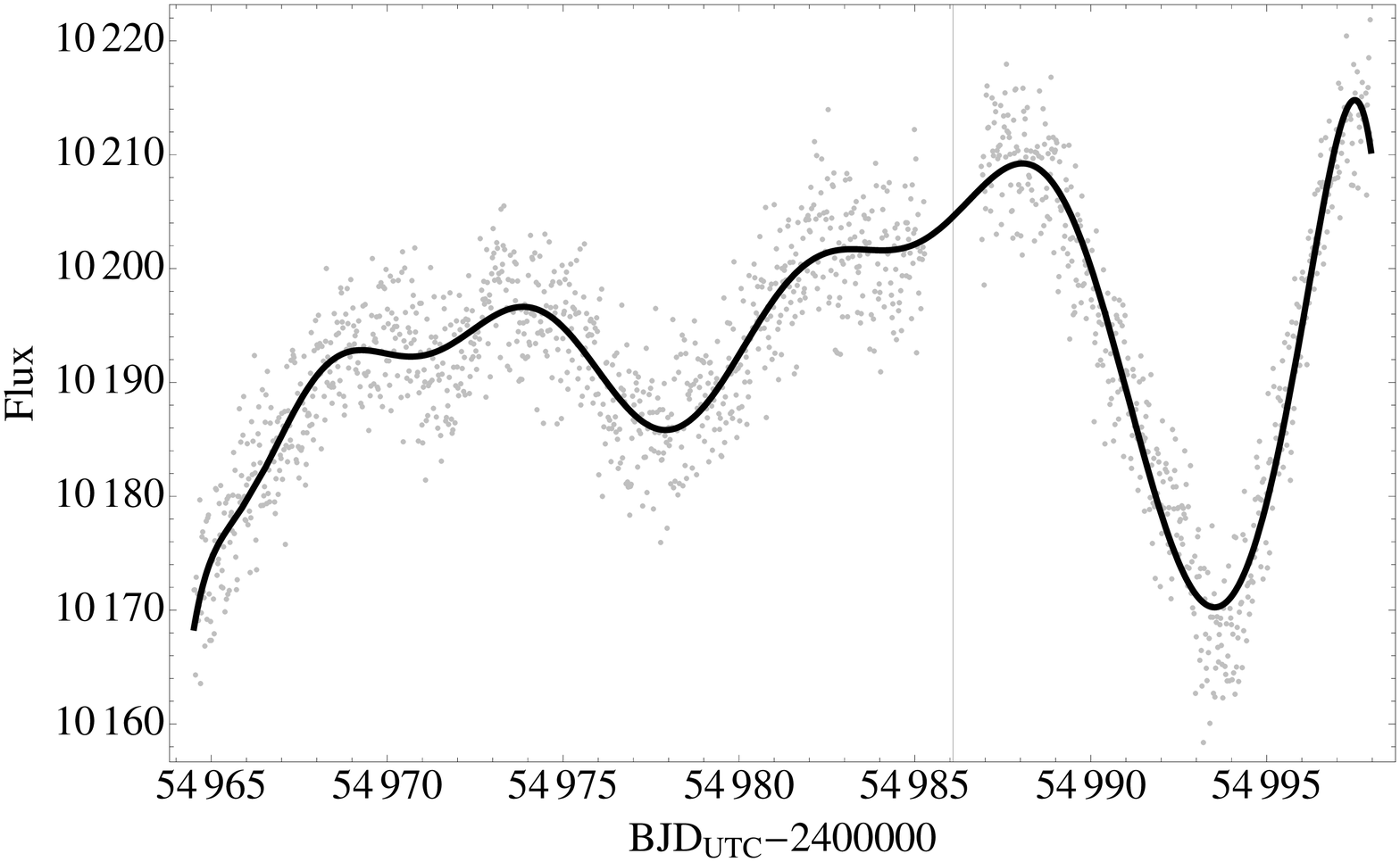,width=75mm}}
\subfigure[Quarter 2
\label{fig:HCV439_Q2raw}]
{\epsfig{file=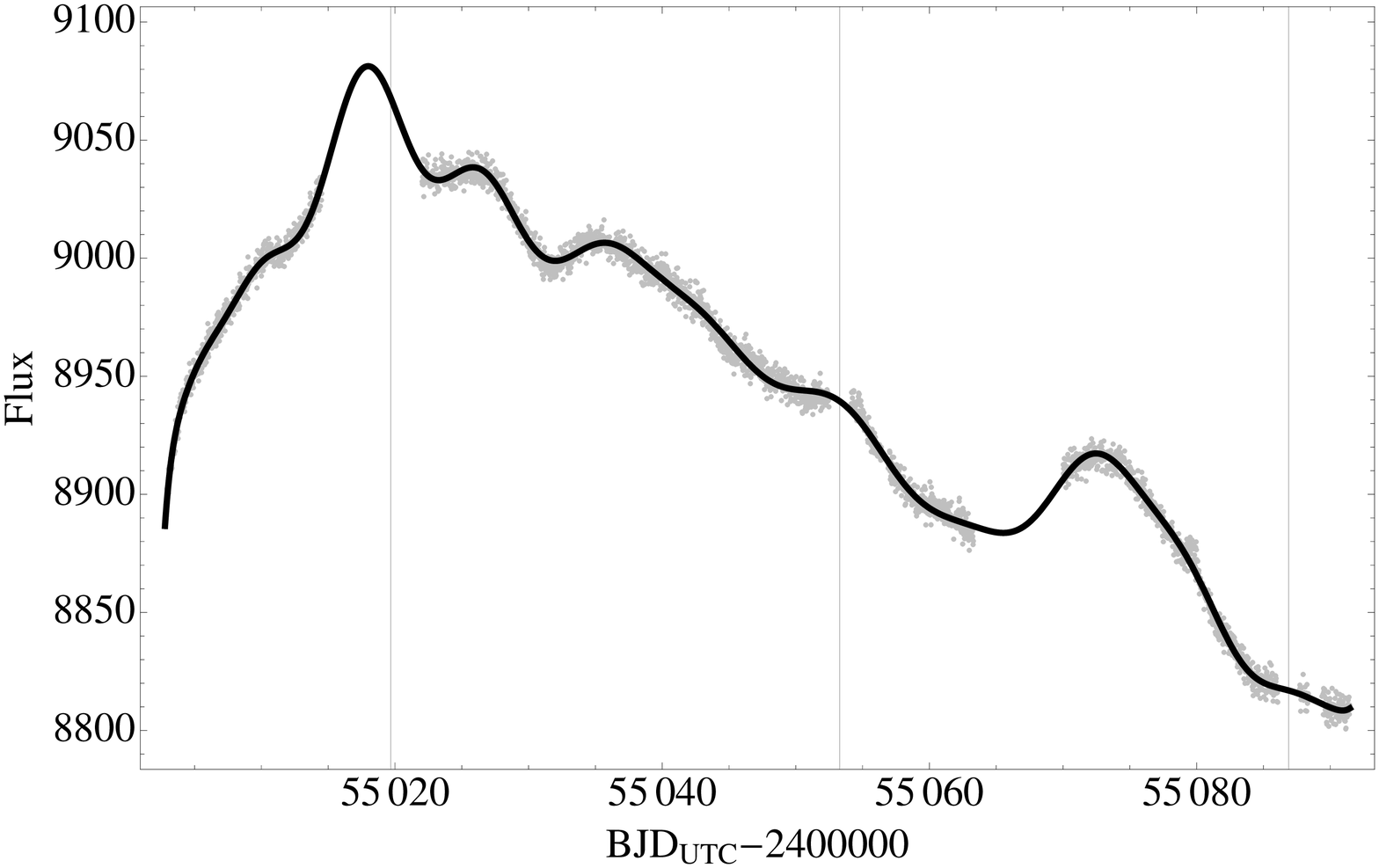,width=75mm}}\\
\subfigure[Quarter 3
\label{fig:HCV439_Q3raw}]
{\epsfig{file=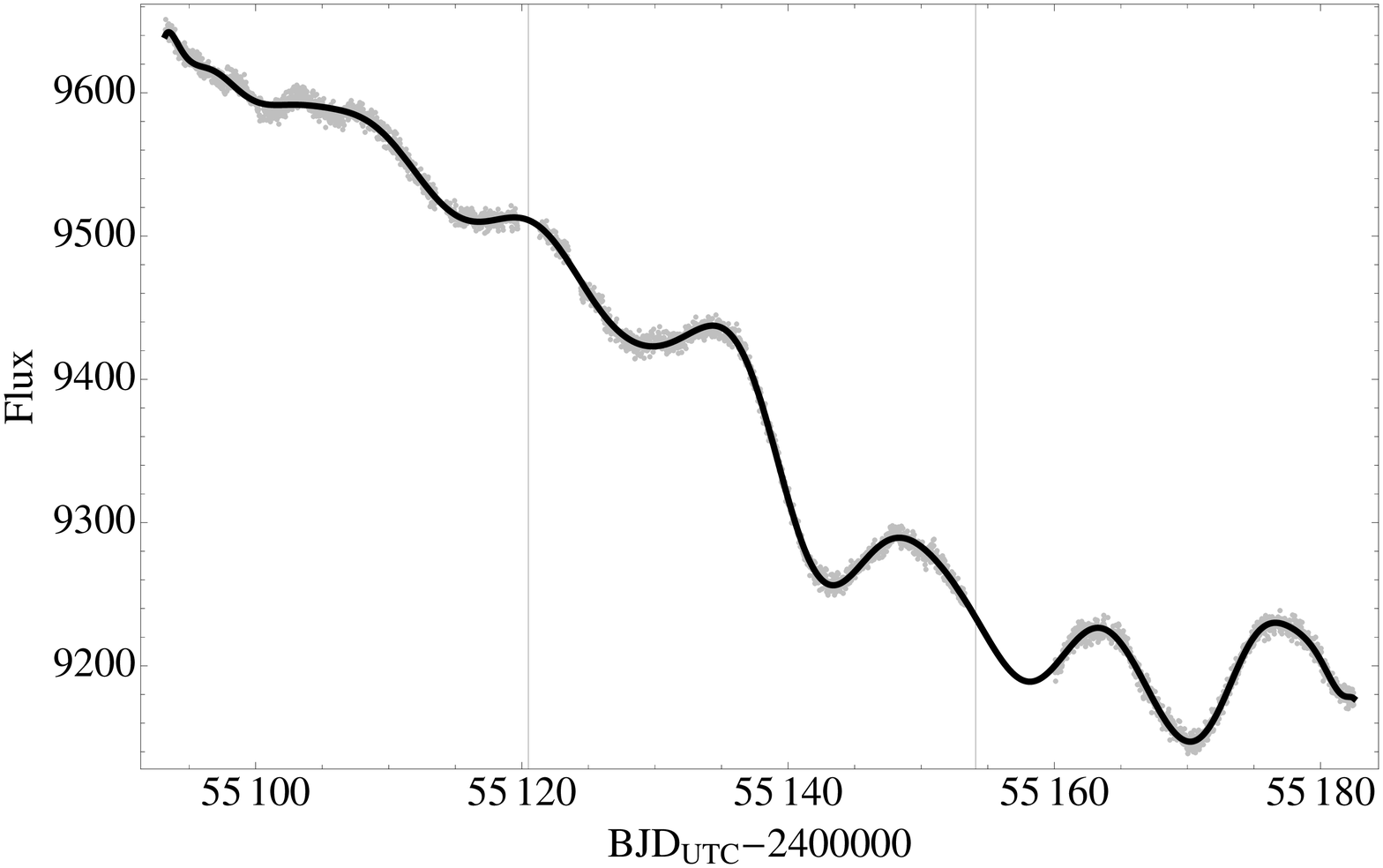,width=75mm}}
\subfigure[Quarter 4
\label{fig:HCV439_Q4raw}]
{\epsfig{file=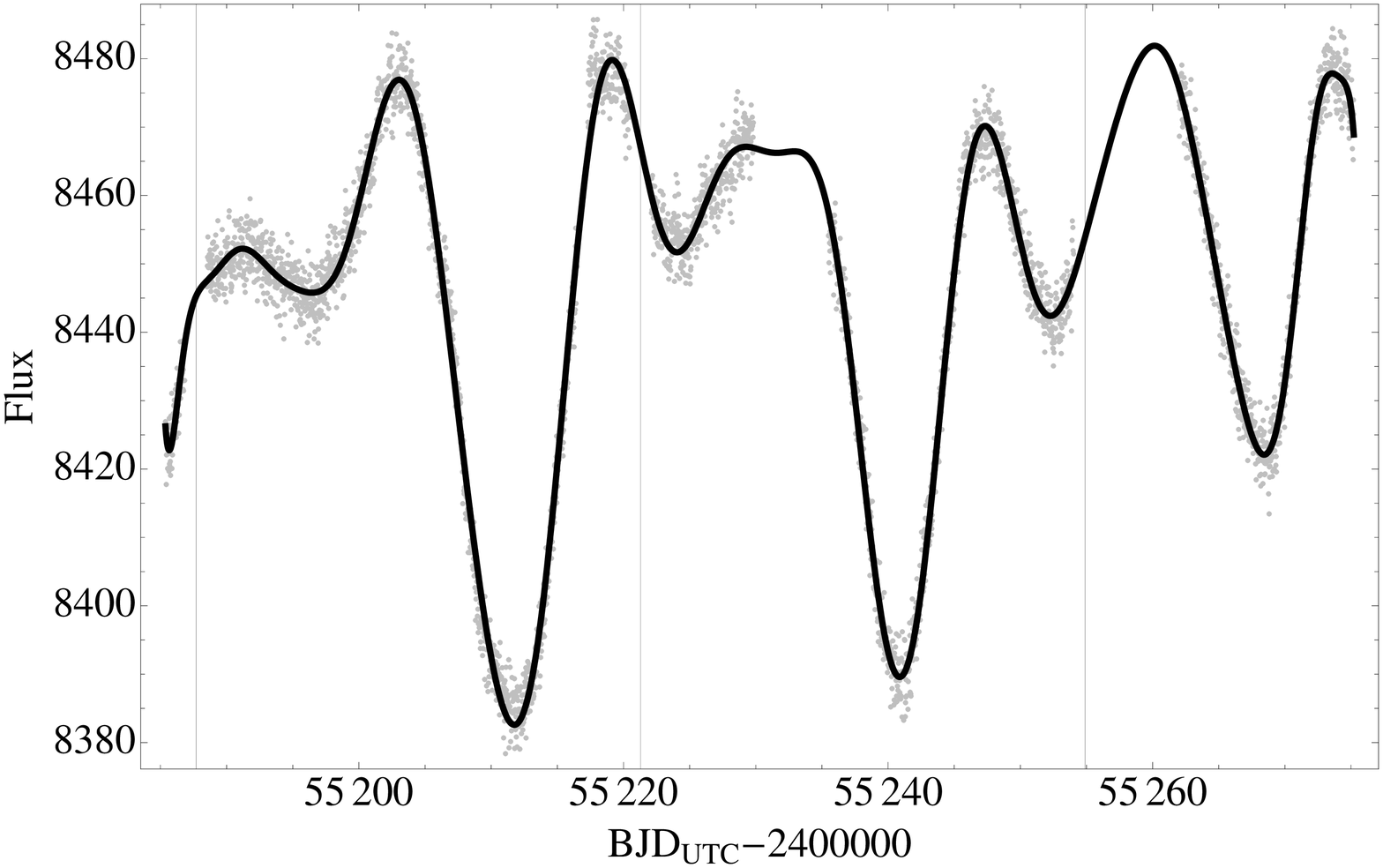,width=75mm}}\\
\subfigure[Quarter 5
\label{fig:HCV439_Q5raw}]
{\epsfig{file=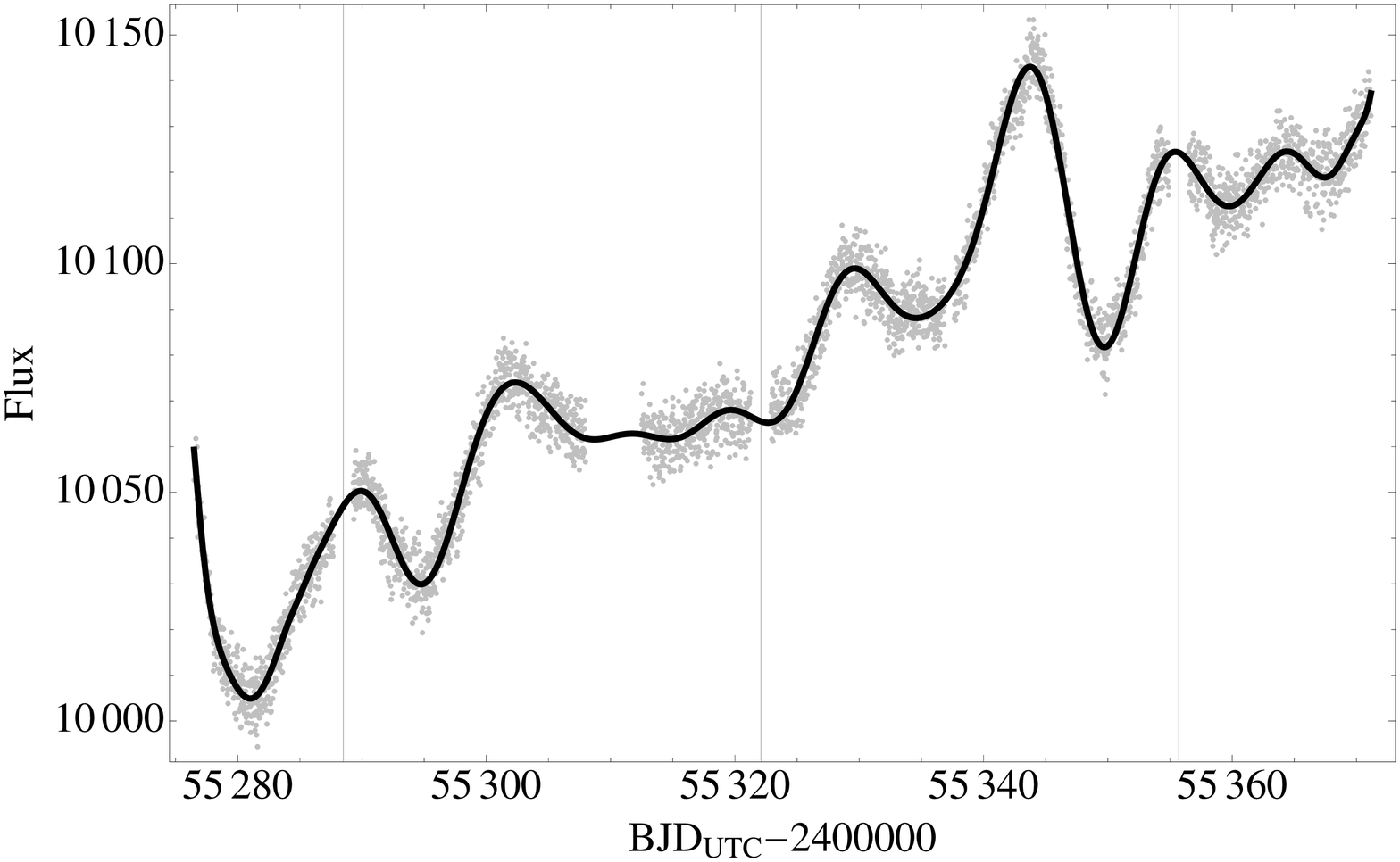,width=75mm}}
\subfigure[Quarter 6
\label{fig:HCV439_Q6raw}]
{\epsfig{file=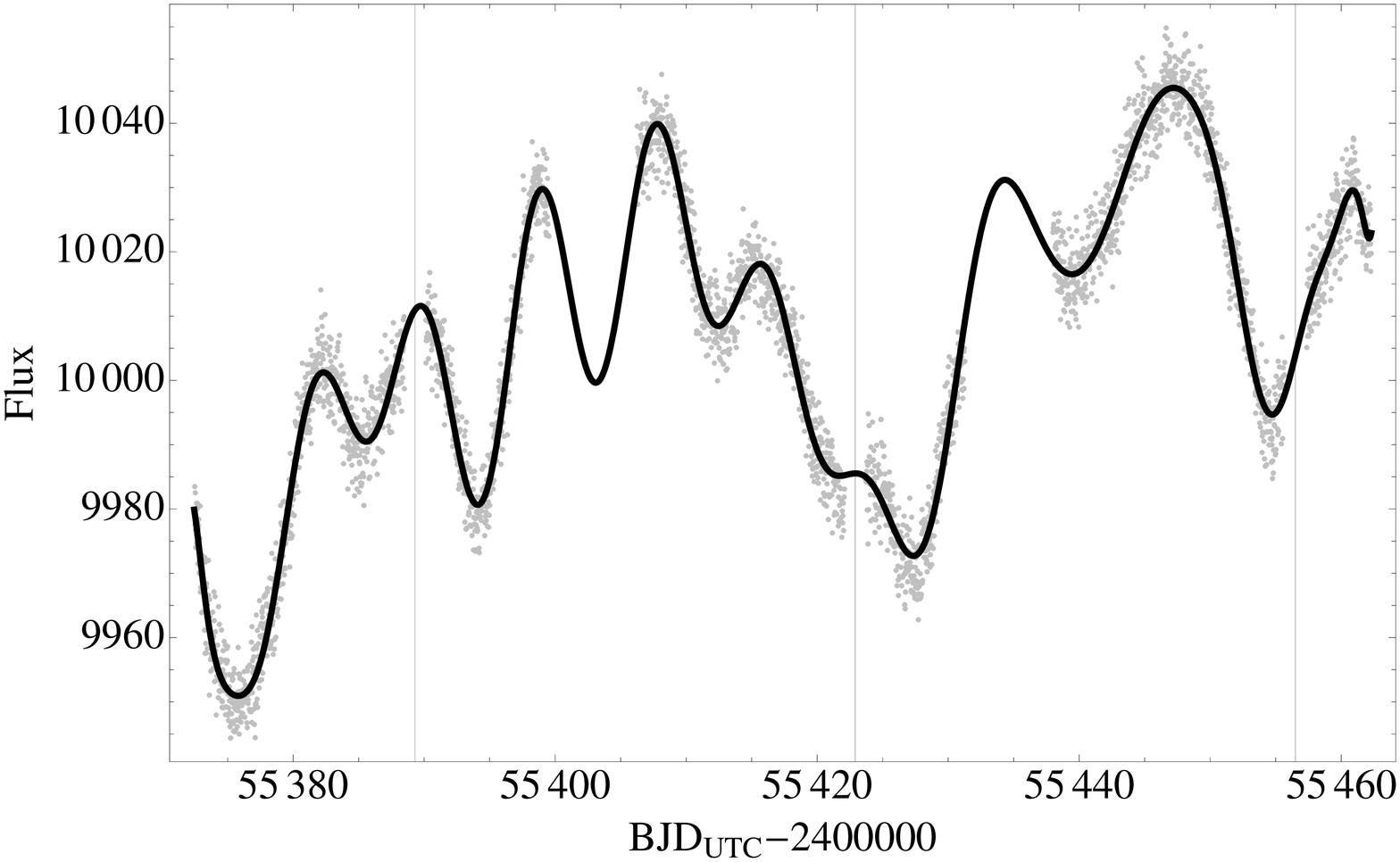,width=75mm}}\\
\caption{
``Raw'' (PA output) flux observed by \emph{Kepler} from DR5
for Q1-6 of the source \koi\ aka \hcv. Overlaid is our 
model for the long-term trend, computed using a discrete cosine transform for 
each data set. Outliers and discontinuous systematic effects have been 
excluded. Transits (removed) marked with vertical gridlines.
\label{fig:HCV439_QAraw}}
\end{figure}

The Q1 to Q6 PA photometry are shown in Fig.~\ref{fig:HCV439_Q1raw}, 
\ref{fig:HCV439_Q2raw}, \ref{fig:HCV439_Q3raw}, \ref{fig:HCV439_Q4raw},
\ref{fig:HCV439_Q5raw} \& \ref{fig:HCV439_Q6raw} respectively. One challenge in 
attempting a correction is assessing which components are astrophysical in 
nature and which are instrumental. The astrophysical signal of interest is
the transit signal and thus we aim to detrend all other effects and preserve
the eclipse. The spurious trends may be removed by applying a high-pass filter 
to the photometry, in a similar way as was used by \cite{mazeh:2010} for 
CoRoT photometry and \cite{kippingbakos:2011a,kippingbakos:2011b}
for \emph{Kepler} photometry. To remove the long-term trends, we thus applied a 
discrete cosine transform \cite{ahmed:1974} adopted to the unevenly 
spaced data.

We first removed the transit events with a margin of $\pm\mathfrak{T}$\,days 
either side of the times of transit minimum, as predicted from the linear 
ephemeris period and epoch reported in \cite{borucki:2011}. The choice 
of the time $\mathfrak{T}$ is provided later. We also remove outliers, 
identified as those points lying 3-$\sigma$ away from a spline-interpolated 
running median of window-size 600\,minutes. Treating each quarter separately, we 
fitted the remaining data with a linear combination of the first N low-frequency 
cosine functions using

\begin{align}
f_i(t_j) = \cos\Bigg[ \Big(\frac{2\pi}{2 B}\Big) t_j i + \phi_i \Bigg],
\label{eqn:cosinefilter}
\end{align}

where $t_j$ is the timing of the $j^{\mathrm{th}}$ measurement, $i=0,N$ in 
integer steps and $N$ is equal to the rounded integer value
of $(2B/4\mathfrak{T}')$ where $B$ is the timespan of the observations and 
$\mathfrak{T}'$ is the timescale we are protecting. Since we are trying to
protect the timescale $\mathfrak{T}$, we set $\mathfrak{T}' = 3\mathfrak{T}$ to
ensure the timescale of interest in minimally distorted\footnote{For Q6, we
were not able to get a good fit to the raw data using $\mathfrak{T}' = 
3\mathfrak{T}$ and so this factor was relaxed to unity.}. Using a
Levenberg-Marquardt algorithm, we then fit for the linear coefficient, $q_i$, 
for each of the cosine functions, so that the fitted model is given by

\begin{align}
\mathfrak{M}(t_j) = \sum_{i=0}^N q_i f_i(t_j).
\label{eqn:trendmodel}
\end{align}

We then subtracted model $\mathfrak{M}$ from the light curve (including the 
transits). The model is shown over the data for the Q1-Q6 photometry in 
Fig.~\ref{fig:HCV439_Q1raw}, \ref{fig:HCV439_Q2raw}, \ref{fig:HCV439_Q3raw}, 
\ref{fig:HCV439_Q4raw}, \ref{fig:HCV439_Q5raw} \& \ref{fig:HCV439_Q6raw}.
The final time series is clipped to be $\pm 2 \mathfrak{T}$ surrounding
the time of transit minimum, as based upon the linear ephemeris of
\cite{borucki:2011}.

{\it Protected timescale.}
The harmonic filter protects a chosen timescale. When searching for
ellipsoidal variations or reflected light, this timescale may be set to be
equal to the period of the transiting planet, as was done in 
\cite{mazeh:2010} and \cite{darkest:2011}. However, if we are 
solely interested in the transit event, then the timescale of interest is much 
shorter than this. Specifically, it is the duration of the transit which is the 
timescale of interest. If we are looking for moons, then auxiliary eclipse 
signals may appear either side of the transit event. Thus, we wish to protect a 
timescale which includes both the transit duration and the surrounding temporal 
coverage corresponding to the region where a moon could reside. Spatially, this 
is the Hill radius and thus we need to define ``a Hill timescale''.

The first-to-fourth transit duration is already known from 
\cite{borucki:2011} to be $T_{14} = 4.3863$\,hours. If this time is 
roughly equal to the time is takes for the planet to traverse two stellar radii, 
$2 R_*$, then the velocity of the planet is given by $v_{P}\simeq2R_*/T_{14}$. 
The Hill timescale, $T_H$, is given by:

\begin{align}
T_H &= \frac{R_H}{v_P}, \nonumber \\
\qquad &\simeq \frac{a_P T_{14}}{2 R_*} \Bigg( \frac{M_P}{3 M_*} \Bigg)^{1/3}.
\end{align}

where $R_H$ is the Hill radius, $a_P$ is the planet's semi-major axis,
$M_P$ is the mass of the planet and $M_*$ is the mass of the star.
In the above, $a_P$, $T_{14}$, $R_*$ and $M_*$ have reasonable estimates
from \cite{borucki:2011}. $M_P$, however, is wholely unknown. Since the 
planet is the Saturn-size regime, we simply adopt a Saturn-like density and
estimate $M_P$ accordingly:

\begin{align}
T_H &\simeq \frac{a_P}{2R_*} \Bigg( \frac{(4/3)\pi R_P^3 \rho_P}{3 M_*} 
\Bigg)^{1/3}.
\end{align}

where $R_P$ and $\rho_P$ are the planet's radius and mean density respectively.
For \hcvb, we estimated $T_H \simeq 0.24$\,days. The protected timescale, 
$\mathfrak{T}$, was then set to be equal to:

\begin{align}
\mathfrak{T} &= 1.2 ( T_{14} + 2 T_H ),
\end{align}

where the 1.2 factor is added as a 20\% buffer. In total, this gives
$\mathfrak{T} = 0.80$\,days as the protected timescale.

{\it Dilution factors.}
In each aperture, a small amount of third light is typically present from the
overlapping point-spread-functions in \emph{Kepler's} relatively crowded field.
To account for this, it is necessary to include the third light in the transit
light curve fits. We follow the method of \cite{kippingtinetti:2010} to
make this correction. Although not available in the MAST headers directly, the 
dilution factors are accounted for in the PDC photometry but not in the PA
photometry. Therefore, we simply take the median of both sets to compute the
dilution factor. This is done for each quarter independently and ranged from
3\% to 9\%.

\paragraph*{Light Curve Analysis}

{\it Star spots.}
The raw transit light curve shows clear evidence for rotating star spots on
the surface of \koi. In particular, Fig.~\ref{fig:HCV439_Q4raw} shows
three deep minima due to a large dark, rotating spot of roughly 10\% the radius
of the star. However, throughout the time series complex spot 
patterns emerge and evolve indicating the behavior is more complex than a 
simple, single spot model. For this reason, a simple periodogram is unable to 
reliably infer a rotation period. A more complex spot-model, accounting for 
multiple spots and differential rotation, is required to interpret the behavior 
of the these events, which is outside of the scope of this work.

For the purposes of this work, the most important consequence of the spots
is that signals which appear like exomoon mutual-events may in fact be star
spot crossings. Equipped with this prior information, a star spot crossing seems 
an a-priori more likely explanation for the TSV signals initially identified. 
These signals also complicate the derivation of upper limits on a putative 
exomoon, as discussed later. Upon downloading the Q4 data in early January, we 
immediately suspected the putative signal was a false-positive. However, we 
continued our analysis because our early model fits revealed that the planet 
exhibited large and complex transit timing variations, as also reported in 
\cite{ford:2011,ford:2012}.

{\it Model fits.}
In fitting the transit light curve, we consider a variety of models to explain
the data. In all cases, the fitted parameter set is the same as that described
in \cite{hek:2012} with two exceptions. Firstly, rather than fitting for
the satellite-to-planet mass ratio ($M_S/M_P$) in planet-with-moon fits, we used 
$(\rho_S)^{2/3}$ i.e. mean density of the satellite to the power of two-thirds.
This was done since a physical density has a better known prior than mass 
ratios. The two-thirds power ensures uniform priors in any derived mass-ratios
since the other density terms feature this index for reasons described in 
\cite{hek:2012}. Secondly, we fit for a photometric noise term, $\sigma_W$,
which is the standard deviation of the noise assuming it to be white. This is
done to propagate the uncertainty of the photometric uncertainties themselves
into our calculation of the Bayesian evidences. $\sigma_W$ is fitted with a 
modified Jeffrey's prior with the inflection point at the median photometric
uncertainty from the DAWG pipeline and the maximum limit being 10 times larger.
We note the best value for this term (from the $\mathcal{M}_T$ fits) is
$\sigma_W = 2.603_{-0.021}^{+0.022}$\,mmag per minute.

In what follows, we assume that the system is a genuine planetary system rather 
than a false-positive, such as a blended eclipsing binary. Details on our 
vetting procedure and blend analysis are provided later. The following
principal models were considered:

\begin{itemize}
\item[{\tiny$\blacksquare$}] $\mathcal{M}_P$ - Planet-only
\item[{\tiny$\blacksquare$}] $\mathcal{M}_T$ - Planet-only with variable times 
of transit minimum (TTV)
\item[{\tiny$\blacksquare$}] $\mathcal{M}_V$ - Planet-only with variable times 
of transit minimum (TTV), transit depths (T$\delta$V), impact parameters and 
$[\rho_{*}^{\mathrm{circ}}]^{2/3}$ values\footnote{$\rho_{*}^{\mathrm{circ}}$
is the light curve derived stellar density assuming a circular orbit.} (thus 
permitting TDV)
\item[{\tiny$\blacksquare$}] $\mathcal{M}_M$ - Planet-with-moon
\item[{\tiny$\blacksquare$}] $\mathcal{M}_{M,R0}$ - Planet-with-moon, defining 
the moon as a point-mass i.e. we fix $R_S = 0$.
\item[{\tiny$\blacksquare$}] $\mathcal{M}_{MT1,M0}$ - Planet-with-moon, 
removing the maximum a-posteriori transit timing variations deduced from 
$\mathcal{M}_T$. Since all TTVs are removed, we must enforce $M_S = 0$.
\item[{\tiny$\blacksquare$}] $\mathcal{M}_{MT2,R0}$ - Planet-with-moon, removing 
the maximum likelihood transit timing variations from a second planet fit and
searching for residual TTVs/TDVs only.
\end{itemize}

The fits were executed using the \multi\ algorithm 
\cite{feroz:2008,feroz:2009}, which is a multimodal nested sampling 
routine \cite{skilling:2004} designed to compute the Bayesian evidence 
in complex parameter space in an efficient manner. For brevity, we direct those
interested to the aforementioned works for further details. \multi\ is coupled 
with the forward-modeling code of \luna\ \cite{luna:2011}, which is 
designed to model the transit light curves of a planet-with-moon accounting for 
mutual events, auxiliary transits, dynamical perturbations and non-linear limb 
darkening in an analytic manner. 

Quadratic limb darkening coefficients were estimated in a similar manner to that 
described in \cite{kippingbakos:2011a}. For this calculation, we assumed 
the effective temperature and surface gravity of the star to be that reported in 
\cite{borucki:2011} ($T_{\mathrm{eff}} = 5127$\,K and 
$\log g_* = 4.59$), which in turn come from the Kepler Input Catalogue (KIC). 
Our later spectroscopic analysis shows these to be excellent estimates. For the 
{\it Kepler} bandpass, we used the high resolution {\it Kepler} transmission 
function found at http://keplergo.arc.nasa.gov/CalibrationResponse.shtml. We 
employed the atmosphere model database from \cite{kurucz:2006} providing 
intensities at 17 emergent angles, which we interpolated linearly at the adopted 
$T_{\mathrm{eff}}$ and $\log g_*$ values. The passband-convolved intensities at 
each of the emergent angles were calculated following the procedure in 
\cite{claret:2000}. This whole process is performed by a Fortran code 
written by I. Ribas. To compute the coefficients we used the limb 
darkening law given in Equation~\ref{eqn:LD}:

\begin{align}
\frac{I_{\mu}}{I_1} &= 1-u_1(1-\mu) - u_2 (1-\mu)^2,
\label{eqn:LD}
\end{align}

where the various terms are defined in \cite{claret:2000}. The final 
coefficients resulted from a least squares singular value decomposition fit to 
11 of the 17 available emergent angles. The reason to eliminate 6 of the angles 
is avoiding excessive weight on the stellar limb by using a uniform sampling 
(10 $\mu$ values from 0.1 to 1, plus $\mu=0.05$), as suggested by 
\cite{diaz:1995}. This leaves us with $u_1 = 0.3542$ and $u_2 = 0.3607$.

\begin{figure*}
\begin{center}
\includegraphics[width=16.8 cm]{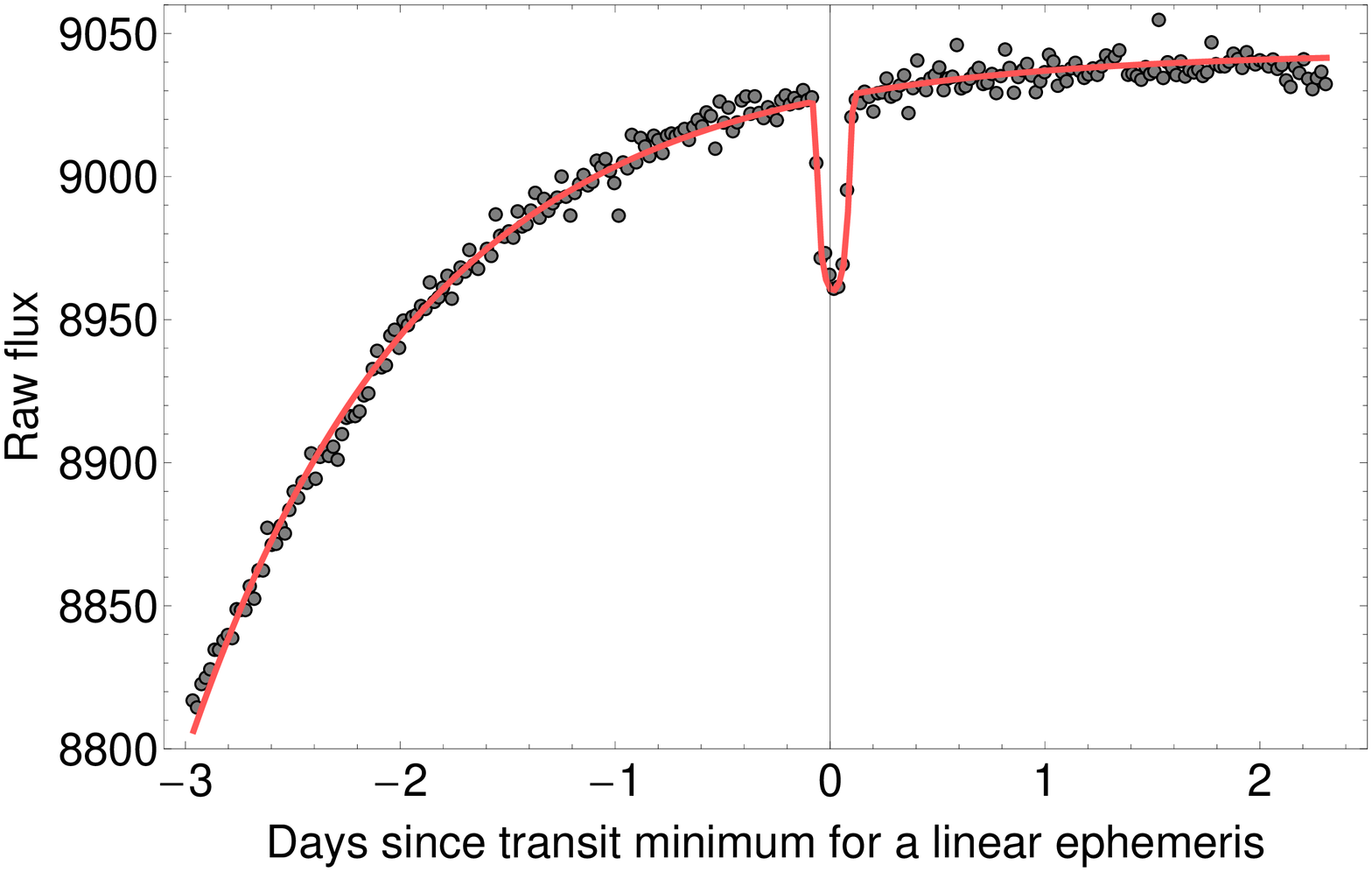}
\caption{Maximum likelihood model (black line) overlaid with the 
long-cadence \emph{Kepler} data for \koi, surrounding the ramp affected transit.
The simple exponential ramp model is shown to provide an excellent description 
of the instrumental effect.}
\label{fig:ramp_fit}
\end{center}
\end{figure*}

{\it Ramp correction.}
During the \emph{Kepler} time series, there are a few safe mode events where the
telescope stopped observing. These result in a pause in the continuous
photometry followed by an exponential ramp as the telescope starts observing
again. This ramp, likely similar to the charge trapping effect seen with
\emph{Spitzer} \cite{agol:2010}, lasts for around one week and we 
usually simply clip the affected data. Unfortunately, the second transit 
observed occurs during one these ramps. In order to maximize the available data, 
we decided to correct for the ramp effect and recover this transit. We apply a 
simple exponential decay model of the form

\begin{align}
F'/F &= a_0 - a_1 \exp(-t/a_2),
\end{align}

where $F'$ is the flux affected by the ramp, $F$ is the flux corrected for
the ramp and $t$ is the time since the start of the ramp. We also tried a double 
exponential, similar that advocated by \cite{agol:2010}, but found the 
two timescales converged to a single value. In fitting the data, we fit the 
parameters $a_0$ (absorbed by the \textbf{OOT} vector in practice), $a_1$ and 
$a_2$ simultaneously to fitting the transit model. We find that this simple model 
provides an excellent description of the ramp effect, as visible in 
Fig~\ref{fig:ramp_fit}. We note that the ramp timescale was 
found to be best modeled by $a_2 = 1.1024_{-0.0069}^{+0.0069}$\,days (from 
$\mathcal{M}_T$ fits), which may be useful to other observers.

{\it Model selection.}
In searching for an exomoon, one must conduct model selection between the 
various hypotheses which could explain the data. We define our null model to be
that of a transiting planet without a moon and with static parameters; model 
$\mathcal{M}_P$. This simply assumes a constant linear ephemeris with a constant
duration and depth every transit.

We also consider models of a planet without a moon, but with perturbations.
The simplest type of perturbation model we consider is that of a planet with
varying times of transit minimum (i.e. a planet experiencing TTV) which we
dub $\mathcal{M}_T$. Model $\mathcal{M}_V$ extends this to allow for variable
depth and duration as well. Models $\mathcal{M}_M$ and $\mathcal{M}_{M,R0}$ are 
the planet-with-moon model, as simulated from \luna, except the latter assumes
a fixed zero-radius moon ($R_S= 0$); i.e. considers TTVs and TDVs only.

Due to the presence of a second planet inducing the TTVs (as discussed in the
main text), we also tried models $\mathcal{M}_{MT1,M0}$ and 
$\mathcal{M}_{MT2,R0}$. The first model removes all transit timing variations by 
subtracting the maximum a-posteriori transit times deduced using $\mathcal{M}_T$ 
from the original data and then refitting for a zero-mass exomoon (we cannot fit 
for a moon mass if we have forcibly removed all TTVs). The second subtracts the 
maximum likelihood model for a second planet (discussed later) causing the TTVs 
and then fits the adjusted data for a planet-with-moon. In this case, we
also fix the moon radius to be zero since we later hypothesize that stellar
activity may be inducing false-positive moon-like eclipses. Therefore, this
model looks for residual TTVs and TDVs only. Because these fits use 
modified data for inputs, they cannot be directly compared to the other models 
and require a custom null-model for comparison in each case.

Model selection is performed by comparing the Bayesian evidence of each model.
The higher the Bayesian evidence the more likely the model is the correct one.
Computing the Bayesian evidence is computationally expensive and particularly
challenging in the high-dimensional space we are faced with. For example,
$\mathcal{M}_T$ involves 38 free parameters. 

In some cases, we found it was not necessary or possible to formally compute the 
absolute value of the Bayesian evidence, $\mathcal{Z}$. For example, as \multi\ 
iteratively searches through higher $\log\mathcal{Z}$ values, the code can be 
stopped if the $\log\mathcal{Z}$ value greatly exceeds the evidence of a 
competing hypotheses. This allows us to place a lower limit on the confidence of 
such a model, as was done for $\mathcal{M}_T$. Here, we found 
$(\log\mathcal{Z}_T - \log\mathcal{Z}_V) \geq (166.1 \pm 0.7)$ (where model
$\mathcal{M}_V$ has the next-best Bayesian evidence), indicating 
$\geq18.1$-$\sigma$ preference for the TTV model over the TTV+TDV+T$\delta$V 
model. To highlight the computational demands, we point out that even this lower 
limit required over 3\,years of equivalent processing time, including over 4 
billion likelihood evaluations, with a 2.1\,GHz AMD Interlagos CPU. We also note 
that the TTV model is preferred over the static model, $\mathcal{M}_P$, at a 
confidence of $\geq43.9$-$\sigma$, which to our knowledge represents the highest 
formal significance for a TTV detection ever reported. For cases where only a
lower limit on $\mathcal{Z}$ is provided, the posteriors were computed by
re-running \multi\ in constant efficiency mode.

Models $\mathcal{M}_M$ and $\mathcal{M}_{M,R0}$ are found to be relatively 
poor fits to the data with $(\log\mathcal{Z}_{M} - \log\mathcal{Z}_{T}) \leq 
-(635.3 \pm 0.4)$ and $(\log\mathcal{Z}_{M,R0} - \log\mathcal{Z}_{T}) \leq 
-(545.8 \pm 0.4)$. This is visually evident by comparing the TTV model for a 
second planet versus that of a moon in Fig.~1. Since $\mathcal{Z}_T \gg 
\mathcal{Z}_P$, there is a very high probability that transit timing variations 
are present. But since a moon model provides a much lower evidence than the TTV 
model (i.e. $\mathcal{Z}_M \ll \mathcal{Z}_T$), then we deduce that i) TTVs are 
present ii) a moon is not responsible.

Although this conclusion tells us the large TTVs are not being caused by a moon,
they do not exclude the presence of a moon either. The possibility of a moon
as an independent source of TTVs is investigated using models 
$\mathcal{M}_{MT1,M0}$ and $\mathcal{M}_{MT2,R0}$. We did not directly compare 
the Bayesian evidence for these models with the others. This is because the 
input data was manipulated in each case by subtracting times of transit minimum 
from each epoch. In order to perform a reliable model comparison, we generated a 
custom null model for each: $\mathcal{M}_{MT1,\mathrm{null}}$ and 
$\mathcal{M}_{MT2,\mathrm{null}}$. These were accomplished by re-running the fit 
with the size and mass of the moon set to zero, allowing us to remove the moon 
terms as free parameters. The results of these fits will be discussed later in a 
dedicated section.

The priors and Bayesian evidence values for each model we attempted are provided
in Tables~\ref{tab:priors}\&\ref{tab:evidences} respectively. Table~1 provides 
the final physical system parameters. Table~\ref{tab:TTVs} provides the final 
system parameters for instrumental terms and times of transit minimum.

\paragraph*{A Search for an Occultation}

Searching for an occultation is challenging due to the large transit timing
variations present. In order to accommodate for this, we allow the occultations
to have their own timing variations. To minimize the number of free parameters,
the $p$, $b_P$, $[\rho_{*}^{\mathrm{circ}}]^{2/3}$, $P_P$ and $\tau_0$ 
parameters were sampled from a Gaussian prior derived from the $\mathcal{M}_T$ 
model fit. Since any eccentricity induced timing offsets should be absorbed by 
the occultation timing variations, fitting for eccentricity terms would be 
essentially fitting redundant parameters and thus we fix $e_P=0$. This left 
us with 15 OOT parameters, 15 occultation times, 1 noise parameter ($\sigma_W$) 
and 1 term for $(\mathcal{F}_P/\mathcal{F}_*)$, the flux-per-unit-area ratio of 
the planet and star. We ran two fits; one where 
$(\mathcal{F}_P/\mathcal{F}_*)=0$ (in such a case the occultation times are not 
required as free parameters) and one where $0<(\mathcal{F}_P/\mathcal{F}_*)<1$ 
and was a uniform prior, allowing us to perform a model comparison later. Note 
that under the assumption that the system is a real planetary system, the TTVs 
suggest $e_P\simeq0$ anyway, as described in the main text.

A comparison of the Bayesian evidence from the null fit versus the occultation
fit yields $(\log\mathcal{Z}_{\mathrm{occ}} - \log\mathcal{Z}_{\mathrm{null}}) 
= (-2.43 \pm 0.44)$. Thus, the null hypothesis of there being no occultation 
present is the preferred model.

Although no occultation is detected, we may use our results to place upper
limits on the occultation depth. The marginalized posterior of of the
occultation depth yields $\delta_{\mathrm{occ}} = 9.9_{-7.4}^{+14.9}$\,ppm,
and places a 3-$\sigma$ upper limit of $\delta_{\mathrm{occ}} < 71.0$\,ppm.
Assuming a geometric albedo of unity ($A_g = 1$), the reflected light component 
of the occultation is expected to be 3.9\,ppm. Assuming the occultation
is due to reflected light only, our occultation depth limit corresponds to
$A_g < 18.0$ i.e. we are unable to constrain the albedo of \koione\ to any
physically plausible range.

Given our insensitivity to reflected light, our upper limit is more robust and
meaningful when interpreted as an upper limit on thermal emission. 
\emph{Kepler}'s visible bandpass is not well-suited to detecting thermal
emission but a meaningful constraint can still be derived. Treating
the star and planet as black bodies and integrating over the custom 
\emph{Kepler} bandpass, we find $T_P < 2442$\,K to 3-$\sigma$ confidence 
(assuming $T_* = 5155$\,K).

\paragraph*{Vetting the Planetary System}

So far, we have assumed the observations are due to an unblended planet
transiting a star. Here, we consider alternative models which do not
require the system to include a planet. To avoid confusion with the earlier
model fits which assume the eclipsing object is an unblended planet, we will
dub these models as hypotheses, $\mathcal{H}_{i}$. There at least three such 
hypotheses which could potentially explain the data:

\begin{itemize}
\item[{\tiny$\blacksquare$}] $\mathcal{H}_{P}$: No blend is present and thus we 
have a planet transiting a star
\item[{\tiny$\blacksquare$}] $\mathcal{H}_{EB,33.6}$: A blended eclipsing binary 
(EB) with the eclipsing bodies on an orbital period of $33.6$\,d
\item[{\tiny$\blacksquare$}] $\mathcal{H}_{EB,67.2}$: A blended eclipsing binary 
(EB) with the eclipsing bodies on an orbital period of $67.2$\,d
\end{itemize}

In the last two cases, the eclipsing binary could be a larger planet (e.g.
Jupiter-sized) eclipsing a star and thus could still be considered a genuine
planetary system. Further, the blend source could be foreground, background, 
associated or a mixture of multiple sources. The last two cases can also be 
considered in two flavors i) EB on a circular orbit ii) EB on an eccentric 
orbit. The possibility of an unblended grazing eclipsing binary is included
in the model $\mathcal{H}_{P}$, and a blended grazing eclipsing binary within
$\mathcal{H}_{EB,33.6}$ and $\mathcal{H}_{EB,67.2}$.

In general, one expects the false-positive rate for Kepler Objects of Interest
to be quite low, with recent estimates arriving at $\lesssim10$\% 
\cite{morton:2011}. Nevertheless, we will here investigate the 
possibility of the blended EB scenarios mimicking a planetary system. We will 
approach this problem using several tools 1) a centroid analysis 2) constraints 
from the spectroscopy 3) model selection with Bayesian evidence determinations 
of the transit light curve shape (a blend analysis) 4) dynamical constraints 
from the timing variations and stability arguments.

\paragraph*{A Centroid Analysis}

{\it Overview.}
The DAWG pipeline output provides flux-weighted centroid positions for all 
observed targets. \cite{batalha:2010} have demonstrated that the very 
small shifts in centroid, expected for blended occulting sources, can be 
measured accurately from the \emph{Kepler} data. Consider two physically 
separated sources with overlapping PSFs. The computed centroid position is 
flux-weighted with respect to these two sources. When one of the sources is 
eclipsed, its flux temporally decreases and thus the flux-weighted centroid 
shifts towards the other source. Therefore, the detection of a shift in 
flux-weighted centroids during the eclipses would indicate the presence of a 
blend source. An example of this technique detecting such a source is for KOI-13 
\cite{mislis:2012}.

{\it Excluded Centroid Shift.}
We extracted the $x$ and $y$ centroid positions for \koi\ surrounding 
$\pm 0.4$\,d of the predicted transit events according to the linear ephemeris
model derived in $\mathcal{M}_P$. We then removed the maximum a-posteriori times 
of transit minimum, $\tau$, for each transit epoch, as computed by the 
$\mathcal{M}_T$ model. This step essentially phases all of the data. Finally, we 
then divided epoch by the median $x$ and $y$ centroid position to remove the 
effect of long-term trends in the centroid positions. No cleaning or detrending 
of the centroids was attempted.

\begin{figure*}
\begin{center}
\includegraphics[width=12 cm]{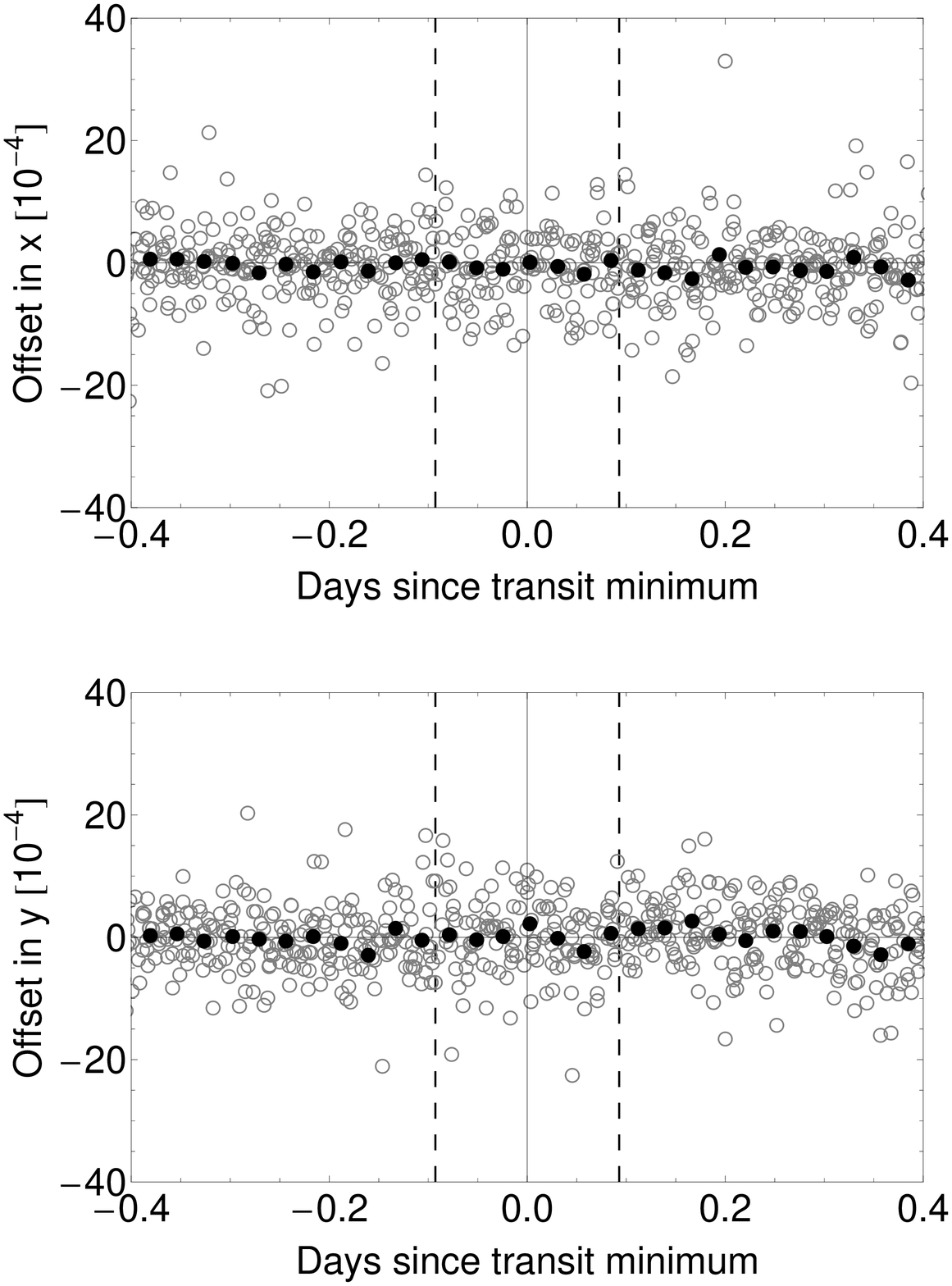}
\caption{$x$ and $y$ centroid positions of \koi\ relative to the median 
value over a 0.8\,day range, surrounding the transits of \koib. Data temporally
offset and phased to account for the transit timing variations. The vertical
grid lines mark the first and fourth contact points. We find no deviation of
the centroids between in- versus out-of-transit, which would have indicated
a separated blend source.
}
\label{fig:centroidplot}
\end{center}
\end{figure*}

The phased centroid positions, shown in Fig.~\ref{fig:centroidplot}, display
no obvious up or down pattern during the transit events (marked by the vertical
gridlines). The scatter can be seen to be comparable, but somewhat larger than,
that derived for KOI-13 in Fig.~2 of \cite{mislis:2012}. Taking the mean
and standard deviation of the in- versus out-of-transit centroid positions we
find:

\begin{align}
\Delta x &= x_{\mathrm{out}} - x_{\mathrm{in}} = (0.09 \pm 0.57) \times 10^{-4}, \\
\Delta y &= y_{\mathrm{out}} - y_{\mathrm{in}} = -(0.07 \pm 0.58) \times 10^{-4}.
\end{align}

Defining $\Delta r = \sqrt{\Delta x^2 + \Delta y^2}$, we determine
$\Delta r = (0.12 \pm 0.57)\times 10^{-4}$, where all units thus far have been 
given in units of pixel position. This analysis clearly indicates that the
data are consistent with no separated blend source present. The 3-$\sigma$
upper limit, converted to arcseconds, corresponds to $\Delta r < 0.68$\,mas,
demonstrating the impressive performance of \emph{Kepler} once again. 

{\it Excluded Blends.}
We here describe a toy model to interpret this upper limit. We consider
two sources of flux $F_*$ and $F_B$ (star and blend source), where the star
is transited. The flux-weighted positions, $r$, of the centroid in- and 
out-of-transit are given by:

\begin{align}
r_{\mathrm{out}} &= \frac{r_* F_* + r_B F_B}{F_* + F_B}, \\
r_{\mathrm{in}} &= \frac{r_* F_* (1-\delta) + r_B F_B}{F_* (1-\delta) + F_B},
\end{align}

where $\delta$ is the unblended eclipse depth. Since we only detect the
blended eclipse depth, $\delta_{\mathrm{obs}}$, we must convert between the two
using the expression from \cite{kippingtinetti:2010}:

\begin{align}
\delta &= \delta_{\mathrm{obs}} (1+\beta),
\end{align}

where $\beta = F_B/F_*$. If we also make the replacement $r_B = r_* + \Delta r$,
one can write:

\begin{align}
\Delta r_{\mathrm{obs}} &= r_{\mathrm{in}} - r_{\mathrm{out}}, \\
\qquad &= \frac{\beta \delta_{\mathrm{obs}} \Delta r}{(1+\beta) (1 - \delta_{\mathrm{obs}})}.
\end{align}

Here, $\Delta r_{\mathrm{obs}}$ represents the observed change in centroid
position during the eclipse, in arcseconds, and $\Delta r$ represents
the physical separation of the two sources on the sky, in arcseconds. Defining
the blend factor as $B=(1+\beta)$, one may solve inverse the above expression to
solve for $B$, as a function of $\Delta r$:

\begin{align}
B(\Delta r) &= \frac{ \delta_{\mathrm{obs}} \Delta r }{ \delta_{\mathrm{obs}} 
\Delta r - \Delta r_{\mathrm{obs}} + \delta_{\mathrm{obs}} 
\Delta r_{\mathrm{obs}} }
\end{align}

\begin{figure*}
\begin{center}
\includegraphics[width=12 cm]{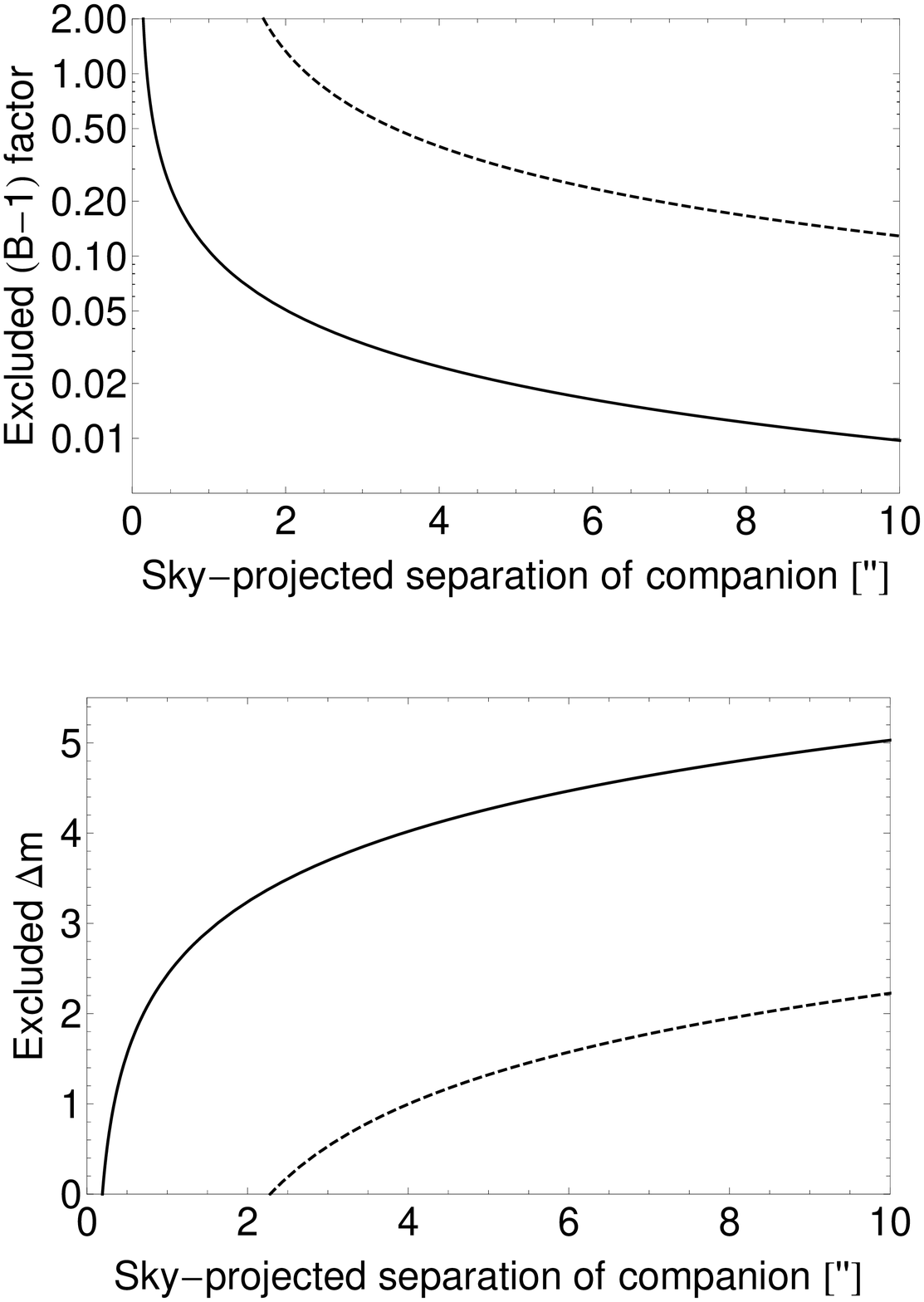}
\caption{By comparing the in-transit to out-of-transit centroid position
for \koib\ (solid) and \koithree\ (dashed), we find no evidence for a shift 
corresponding to a separated blend source. Our upper limits allow us exclude the 
blend factor, $B$, as a function of the separation of a companion. The top panel 
shows the results when plotted for $B$, the lower panel when plotted for 
magnitude difference.
}
\label{fig:excludedblends}
\end{center}
\end{figure*}

Using this simple model, we plot the excluded values of $B$ as a function of
$\Delta r$ in Fig.~\ref{fig:excludedblends}. Note that this result is
purely based on the centroid shifts. To quote several values from the figure,
$B<1.239$ for $\Delta r<0.5''$, $B<1.051$ for $\Delta r<2''$ and
$B<1.016$ for $\Delta r<6''$. A blend factor of $\simeq$10\% does not impact
significantly on our results and thus only blends within $\Delta r<0.5''$
could possibly cause a planet false-positive. We note that, in general, such
a closely-space companion is quite rare, with a recent adaptive optics
campaign on suitable KOIs finding only 6.7\% of KOIs have a companion within
0.5''.

Fig.~\ref{fig:excludedblends} also provides the same constraints as computed
for the inner transiting planet candidate \koithree\ (dashed line). Details
on the detection of this candidate are provided later. Due to the much 
smaller depth, the upper limits are less constraining but nevertheless still
consistent with the absence of a blend source.

\paragraph*{A Spectroscopic Analysis.}

{\it Observations.}
Spectroscopic observations of the star \hcv\ were carried out with the 
Astrophysical Research Consortium Echelle Spectrograph (ARCES) on the Apache 
Point Observatory 3.5\,m telescope located at Apache Point Observatory (APO) in 
New Mexico. We used a $1\farcs 6 \times 3\farcs2$ slit which delivers a spectral 
resolution of $\Delta \lambda /\lambda \approx 31,\!000$ over a spectral range 
of 3200--$10,\!000$\,\AA. We obtained a total of two thirty minutes exposures on 
the star \hcv\ on the night of 2012 Jan 31 which, when combined, yielded a total 
S/N of 11 at $5100\,\AA$. After performing a standard overscan correction, we 
removed cosmic rays, extracted the spectra, applied a flat-field correction, and 
determined the dispersion correction from a ThAr lamp spectrum using standard
techniques in the IRAF's IMRED, CRUTIL, and ECHELLE packages.

{\it Stellar Parameters.}
We used the Yonsei-Yale (YY) isochrones \cite{yi:2001} to determine the 
physical properties of the host star. The first input into this analysis were 
the stellar atmosphere parameters determined from the APO 3.5\,m spectra. We 
used the Stellar Parameter Classification (SPC) method 
to derive the stellar atmosphere parameters. SPC cross-correlates the observed 
spectrum against a grid of synthetic spectra drawn from a library calculated by 
John Laird using Kurucz models \cite{kurucz:1992}. The synthetic spectra cover a 
window of 300\,\AA\ centered near the gravity-sensitive Mgb features and has a 
spacing of 250\,K in effective temperature, 0.5\,dex in gravity, 0.5\,dex in 
metallicity and $1$\,km\,s$^{-1}$ in rotational velocity. To derive the precise 
stellar parameters between the grid points, the normalized cross-correlation 
peaks were fitted with a three dimensional polynomial as a function of effective 
temperature, surface gravity and metallicity. This procedure was carried out for 
different rotational velocities and the final stellar parameters were determined 
by a weighted mean of the values from the spectral orders covered by the 
library.

The isochrone analysis made used of the stellar effective temperature 
$T_{\mathrm{eff}} = (5155 \pm 105)\,K$, and the metallicity 
$[\mathrm{Fe}/\mathrm{H}] = (0.41 \pm 0.10)$ from the SPC analysis. The second 
input into the YY analysis came from the light curve analysis of the 
{\it Kepler} data. The transit duration is closely related to the $a_P/R_*$ 
parameter, which in turn determines the mean stellar density $\rho_*$ 
\cite{seager:2003}. In general, this trick is only possible for systems where 
the orbital eccentricity is precisely known \cite{investigations:2010}. It will 
be shown in the dynamical analysis discussion later that planet b's eccentricity 
is indeed strongly constrained from the TTVs to be near-circular. We proceed by 
using the light curve derived stellar density posterior from our model fits, 
giving $\rho_* = 1520_{-170}^{+220}$\,kg\,m$^{-3}$.

\begin{figure}[h!]
\includegraphics[width=14 cm]{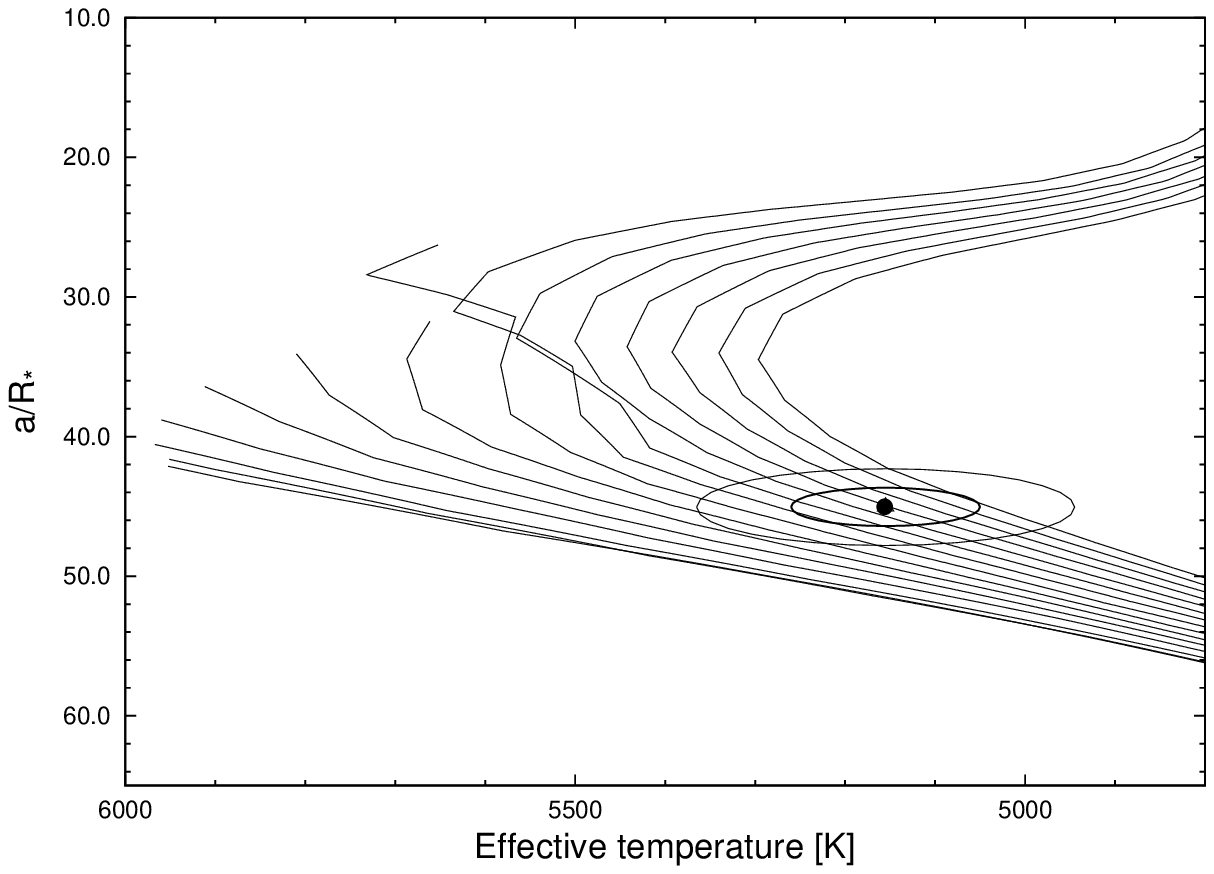}
\caption{YY-isochrone analysis of \koi. Using our spectrosopic observations
and the constraint on $\rho_*$ from the transit light curve of \koib, we are
able to determine precise parameters for the host star. The backdrop of 
isochrones is for ages 0.2, 0.5, 1.0, 2.0, \ldots 13.0\,Gyr (from bottom to top)
and $[\mathrm{Fe}/\mathrm{H}] = 0.41$. The solution indicates an old star 
with $10\pm3$\,Gyr age.}
\label{fig:isochrones}
\end{figure}

The mean stellar density acts as a luminosity indicator for the star; smaller 
density typically means more evolved stars.  Another possible luminosity 
indicator would be the surface gravity of the star, as determined from the
spectroscopic analysis. If the eccentricity of the system is well determined 
and the light curve is of high quality (both of which are true here), then 
$a_P/R_*$ and the corresponding $\rho_*$ is typically a better luminosity 
indicator than $\log g_*$, in the sense that the derived physical parameters 
have a smaller error.  We have generated over 10,000 values of
$T_{\mathrm{eff}}$, $[\mathrm{Fe}/\mathrm{H}]$ and $a_P/R_*$ using their a 
posteriori distribution (assuming Gaussian distribution for the first two), and
searched the YY isochrones for each materialization. About 95\% of
the input values had a matching isochrone. Stellar parameters were
then determined as the median of the resulting distribution. The final
parameters were $M_* = 0.90\pm0.04\,M_{\odot}$, $R_* = 0.94\pm0.04\,R_{\odot}$,
and $\log g_* = 4.44\pm0.04$ (cgs). The isochrones and the final
solution are shown in Fig.~\ref{fig:isochrones}, where the backdrop of 
isochrones is for ages 0.2, 0.5, 1.0, 2.0, \ldots 13.0\,Gyr (from bottom to top)
and $[\mathrm{Fe}/\mathrm{H}] = 0.41$. The solution indicates an old star 
with $10\pm3$\,Gyr age.

In regard to vetting of the system, we find no evidence for double-lines 
indicating a blend. We also find the spectral classification of the star
very well described as a dwarf main-sequence star, and exclude the possibilities
such as a giant star or white dwarf. We note that our classification is in close 
agreement with the KIC determination reported in \cite{borucki:2011}.

\paragraph*{A Blend Analysis}

{\it Overview.}
The third vetting tool we use is model selection with Bayesian evidence 
determinations of the transit light curve shape i.e. a blend analysis. Blend 
analyses have become a powerful instrument in the toolbox of the \emph{Kepler} 
team too, using their custom BLENDER software \cite{torres:2004,fressin:2011}. 
Here, the team simulate a grid of billions of possible false-positive scenarios 
and compute the odds ratio of valid planet solutions versus valid false-positive 
solutions. Systems strongly favoring the planet solution are considered to be 
``validated''. BLENDER makes use of additional information such as centroid 
positioning, priors of the frequency of eclipsing binaries and multi-color light 
curves (often from \emph{Spitzer} \cite{fressin:2011}). In this work, we limit 
our analysis to purely an inspection of the shape of the light curve. This is 
done to reduce the computational demands of simulating billions of 
false-positives, and yet take advantage of the fact we have other more powerful 
constraints from the dynamics (as discussed later). Further, a centroid analysis 
has already indicated that a blend must be within 0.5'' to have a significant 
impact on our results and thus can be treated as a separate line of evidence.

Although our blend analysis is more simplified than BLENDER, it does take
advantage of the Bayesian evidence for model selection, which is currently
not implemented in BLENDER or other blend analyses in the current literature. 
Our approach is to consider a null hypothesis and then more elaborate models
involving blend scenarios and compute the Bayesian evidence of each model.
Since the more elaborate models include more parameters (i.e. a greater
prior volume), they are penalized in the computation of the Bayesian evidence.
If a blend model provides a Bayesian evidence significantly greater than that of 
the null model ($\Delta\log\mathcal{Z}\gtrsim5$), it passes the first test to 
becoming the preferred model. The second test we impose is that the parameter 
posteriors of the model must correspond to a physically plausible scenario. If 
both of these criteria are satisfied, then the null model can be displaced.

Another possibility is that two or more hypotheses yield approximately equal
Bayesian evidences, such that there is no statistically significant preference
between them. In such a case, we continue to consider these hypotheses as
plausible descriptions of the system and test whether they are consistent with
our subsequent dynamical analysis too.

The null hypothesis is the simplest model which can explain the data and so 
involves the fewest parameters to describe the system. In our case, this
represents just two eclipsing objects without a blend. Whilst an unblended
grazing eclipsing binary could fall into this category, it will be shown shortly
that such a hypothesis is highly improbable. Therefore, the null hypothesis 
essentially represents a planet-sized object transiting a star.

{\it Implementation.}
In our blend analysis, we are only investigating the shape of the light curve.
The dynamical constraints from the timing variations will be discussed later,
and so here we eliminate them by subtracting any TTVs away from a linear
ephemeris. This is accomplished using the maximum a-posteriori transit times
from the model fit $\mathcal{M}_T$ performed earlier. This reduces the number of 
free parameters by 15 and makes the fits far easier to handle computationally. 
However, we still fit each eclipse epoch with a unique baseline to remove any 
residual DC power from the detrending procedure.

For the null hypothesis, $\mathcal{H}_{\mathrm{P}}$, we have 15 OOT baseline
parameters, $p$, $b_P$, $[\rho_{*}^{\mathrm{circ}}]^{2/3}$, $P_P$, $\tau_0$,
2 instrumental terms to describe the ramp effect for the one affected event and
two limb darkening parameters ($u_1$ and $(u_1+u_2)$), giving 24 parameters in 
total. Limb darkening was fitted for to provide a fair comparison to the blend
models where limb darkening cannot be assumed to be the same as the theoretical
models used in the planetary fits. The priors on the limb darkening terms were 
selected such that the brightness profile is positive everywhere and 
monotonically decreasing from limb to center, specifically $0<u_1<2$ and
$0<(u_1+u_2)<1$ \cite{carter:2009}.

We also extended the prior on $p$ to the range $0<p<1$ and similarly for the 
impact parameter $0<b_P<2$. The other fitted terms had the same priors as used 
before (see Table~\ref{tab:priors}).

For the 33.6\,d period EB scenarios, we simply add a single additional term,
a blending factor $B$. Our model for the diluted light curve follows the
prescription of \cite{kippingtinetti:2010}, where $B=1$ indicates no 
blend and $B>1$ indicates a blend. The model is general in that the source of 
the blend could be foreground, background, associated or a mixture. The prior on
the blending factor was chosen to be $1<B<100$, and all other priors were left
unchanged from $\mathcal{H}_P$. In general, one expects the 33.6\,d blended EB
scenario to give an equally good fit to the data as a planet, since the $B$
term is degenerate with the other fitting parameters \cite{kippingtinetti:2010}. 
For the eccentric cases, we fitted for $e_P$ and $\omega_P$ directly using 
$0<\omega_P<2\pi$ and $0<e_P<0.9$ uniform priors. The eccentricity was cut-off 
from extremely eccentric orbits to save CPU time, since solving Kepler's 
equation takes dramatically longer in the extreme eccentricity regime.

The final scenario of a 67.2\,d period blended EB was treated by first modifying
the prior on the period to be uniform around $\pm1$\,day of 67.2\,d. Then, we 
instructed the code to treat the occultations has having zero-limb darkening.
Whilst the ingress/egress of the occultation may in fact have limb darkening, 
the 2nd-to-3rd contact of the transit, which dominates the signal-to-noise, 
cannot have a limb darkened profile. Therefore this approximation contains the 
most important physics of the problem. The primary transit is treated as a limb 
darkened event as before. The occultation depth is equal to 
$p^2 (\mathcal{F}_P/\mathcal{F}_*)$, where $\mathcal{F}$ is the 
flux-per-unit-area of a body. This flux ratio is the only new parameter 
required, for which we use $0<(\mathcal{F}_P/\mathcal{F}_*)<2$ as our prior. In 
general, one expects this scenario to be easier to distinguish against a planet 
due to the lack of curvature in alternate eclipses.

{\it Results: An Unblended Grazing EB.}
The easiest scenario to disregard is that of an unblended grazing eclipsing 
binary. Such a scenario would be permitted in model $\mathcal{H}_{P}$ and may be 
tested for by evaluating the number of posterior samples which satisfy 
$b_P>(1-p)$; the definition of a grazing event. We find that $(b_P+p)<1$ to 
$>99.99$\% confidence and not a single posterior sample landed in this regime. 
Therefore, the hypothesis of an unblended grazing EB is highly improbable.

{\it Results: A Blended Grazing EB.}
A blended grazing eclipsing binary could feature in all four of the alternative
hypotheses; $\mathcal{H}_{EB,33.6}^{c}$, $\mathcal{H}_{EB,33.6}^{e}$,
$\mathcal{H}_{EB,67.2}^{c}$ \& $\mathcal{H}_{EB,67.2}^{e}$. The first thing to
note is that Bayesian evidence of all of these models is not significantly
improved over the null hypothesis, $\mathcal{H}_P$.
In each of the four models, we checked the posterior samples satisfying 
a non-grazing configuration and found $(b_P+p)<1$ to $>99.99$\% in all four.
Therefore, the hypothesis of a blended grazing EB is highly improbable.

{\it Results: A Blended 67.2\,d EB.}
As discussed earlier, the 67.2\,d period blended EB causes a limb darkened
transit but a flat-bottomed occultation. This difference in transit profile
is expected to make the scenario easier to distinguish than the 33.6\,d blended
EB. Indeed, this is what we found. Hypotheses $\mathcal{H}_{EB,67.2}^c$
and $\mathcal{H}_{EB,67.2}^e$ yield $\Delta\log\mathcal{Z}=(-22.7\pm0.5)$ 
and $(-133.1\pm0.5)$ respectively, relative to the null hypothesis 
$\mathcal{H}_P$. This constitutes a $6.4$-$\sigma$ and $16.1$-$\sigma$ 
preference for the null hypothesis, for the two alternative hypotheses 
respectively. Therefore, the hypothesis of a blended 67.2\,d EB is highly
improbable.

{\it Results: A Blended 33.6\,d Circular Orbit EB.}
With all previous scenarios now rejected, we are left with the blended 33.6\,d
EB only, which comes into two flavors: $\mathcal{H}_{EB,33.6}^c$ \& 
$\mathcal{H}_{EB,33.6}^e$. Let us here consider the circular case first. 
Strictly considering the Bayesian evidence results of our light curve profile 
analysis, there is no significant preference between the hypothesis of an 
unblended planetary transit and a blended 33.6\,d EB. Note that these remaining 
blend scenarios include a Jupiter-sized planet transiting a blended 
main-sequence star.

Scenarios involving a giant star or white dwarf can be excluded based upon the 
spectroscopic analysis discussed earlier. The ratio-of-radii in both 
$\mathcal{H}_{EB,33.6}^c$ is constrained to be $p<0.30$ to $>99.99$\%. This 
means we must be dealing with a main-sequence star being transmitted by a 
smaller object, albeit with the possible presence of a blend. 

We discussed earlier how there is no detectable occultation in the data between
the 33.6\,d transits. Under the hypothesis of $\mathcal{H}_{EB,33.6}^c$, the
other object must have $(\mathcal{F}_P/\mathcal{F}_*) < 0.0089$ to 3-$\sigma$
confidence. This indicates a small ($p<0.3$), cool ($T\lesssim 3000$\,K) object
consistent with a planet or brown dwarf.

{\it Results: A Blended 33.6\,d Eccentric Orbit EB.}
The final case of the eccentric 33.6\,d EB yields a slightly improved Bayesian 
evidence over the planet-only model, with a significance of 
$(2.4\pm0.2)$-$\sigma$. We do not consider this significant enough to overturn
the planet hypothesis but nevertheless the scenario is considered plausible
at this stage. A possible reason for the slight improvement is the ability of
an eccentric orbit to generate a more diverse range of limb darkening profiles
than the circular orbit case, when both models utilize quadratic limb darkening.

{\it Results: Conclusions.}
We conclude that the only models which can adequately explain the spectroscopic
analysis and the blend analysis are that of a planet transiting a star 
($\mathcal{H}_P$), a blended 33.6\,d EB on a circular orbit 
($\mathcal{H}_{EB,33.6}^{c}$) and a blended 33.6\,d EB on an eccentric orbit
($\mathcal{H}_{EB,33.6}^{e}$). In all cases a grazing transit configuration is 
excluded.

\paragraph*{A Dynamical Analysis}

The spectroscopic and blend analyses thus far leave us with three hypotheses. 
Here we will evaluate these hypotheses in light of the dynamical constraints 
from both the transit timing variations (TTV) and stability arguments.

It has been established that the observed large transit timing variations 
cannot be caused by a moon. It is shown in the main text that the only plausible
source for such large TTVs is a third body in the system orbiting close to
the 5:3 orbital resonance. These fits do not allow us to directly measure
the mass for the transiting body but do allow us to constrain the
eccentricity to a high degree of confidence. Thus, the first hypothesis we 
may consider is the eccentric 33.6\,d blended EB scenario, 
$\mathcal{H}_{EB,33.6}^{e}$.

{\it Eccentricity Constraints.}
The spectroscopic analysis can be used to provide a mass and radius for the
star using stellar evolution models, as was done earlier. However, if a 
substantial amount of blended light is present then the inferred properties 
would be unreliable. This is particularly salient in light of the faintness of 
the target and the subsequent lower-than-normal SNR spectra obtained. However, 
even in the case of a substantial amount of blended light, we are confident that 
the star is a dwarf on or near the main-sequence. We consider a wide range of
corresponding plausible stellar masses to be 
$0.1\,M_{\odot} \leq M_* \leq 10.0\,M_{\odot}$.

Using this mass range, we re-fit the TTVs each time varying all of the system
parameters. After finding the maximum likelihood solution, we perturb the
parameters in order to derive an upper limit on the orbital eccentricity of
the transiting planet. Across the full stellar range, we find $e_{P} < 0.02$
to 99.9\% confidence. Note that the derived TTVs are insensitive to any
blended light and thus this limit is robust for $B>1$.

We conclude that the hypothesis of an eccentric 33.6\,d blended EB is highly
improbable. This now leaves only two remaining hypotheses to explain the
data: $\mathcal{H}_P$ and $\mathcal{H}_{EB,33.6}^{c}$.

{\it Mass Constraints.}
The two surviving hypotheses are identical in that they both have a body on a 
33.6\,d circular orbit eclipsing a main-sequence dwarf star on a non-grazing 
transit. The only difference is the amount of blended light ($B=1$ versus 
$B>1$). In both cases, the eclipsing object is small ($p<0.3$ to 99.99\% 
confidence) and cool ($\mathcal{F}_P/\mathcal{F}_*<0.0089$ to 3-$\sigma$
confidence, which is robust for any $B$ value). The only plausible blend 
scenarios which could reproduce all of these constraints is a very cool M-dwarf, 
a brown dwarf or a larger planet. Clearly the planetary nature of the transiting 
body is not yet validated as these objects span a wide range of possible masses.

The final and most powerful tool at our disposal is that the TTVs allow us
to measure $M_c/M_*$ for the third body. For a given $M_*$ then, two of the
three masses in the system are known, along with their periods, semi-major
axes, eccentricities and mutual inclination. The only unknown is the mass of
the transiting object. We may place an upper limit on this value by iteratively
increasing the mass until the system becomes dynamically unstable.

Dynamical stability was investigated using the symplectic N-body code known as 
SyMBA \cite{duncan:1998}, simulating the system for 1\,Gyr using an 
integration timestep of 1.5\,d. One can account for the possibility that the 
stellar mass derived from the spectroscopy is unreliable (due to blending) by 
investigating a wider, but plausible, range of stellar masses. To this end, we 
scanned the range $0.8\,M_{\odot} \leq M_* \leq 1.2\,M_{\odot}$. In order to be 
stable for 1\,Gyr, we estimate $M_P/M_* < 5\,M_J$, $<7\,M_J$ and $<9\,M_J$ for 
0.8\,$M_{\odot}$, 1.0\,$M_{\odot}$ and 1.2\,$M_{\odot}$ respectively, all of
which exhibit a very sharp stability boundary. For our best-fit stellar mass
of $0.9$\,$M_{\odot}$ (see earlier spectroscopy discussion), the mass of
\koione\ is constrained to be $M_P<6$\,$M_J$.

We therefore conclude that the transiting object \koione\ must be planetary in
nature (since $M_P<11$\,$M_J$; \cite{spiegel:2011}) and thus refer to the object 
as \koib\ from here-on-in. \koic\ is validated as a planet based upon the 
precise measurement of its mass from the TTVs, specifically 
$M_c/M_* = 3.97_{-0.14}^{+0.17}\times10^{-4}$, corresponding to 
$M_c = 0.376_{-0.020}^{+0.023}$\,$M_J$.

\paragraph*{A Search for Additional Transits}

\begin{figure}[h!]
\includegraphics[width=13.5 cm]{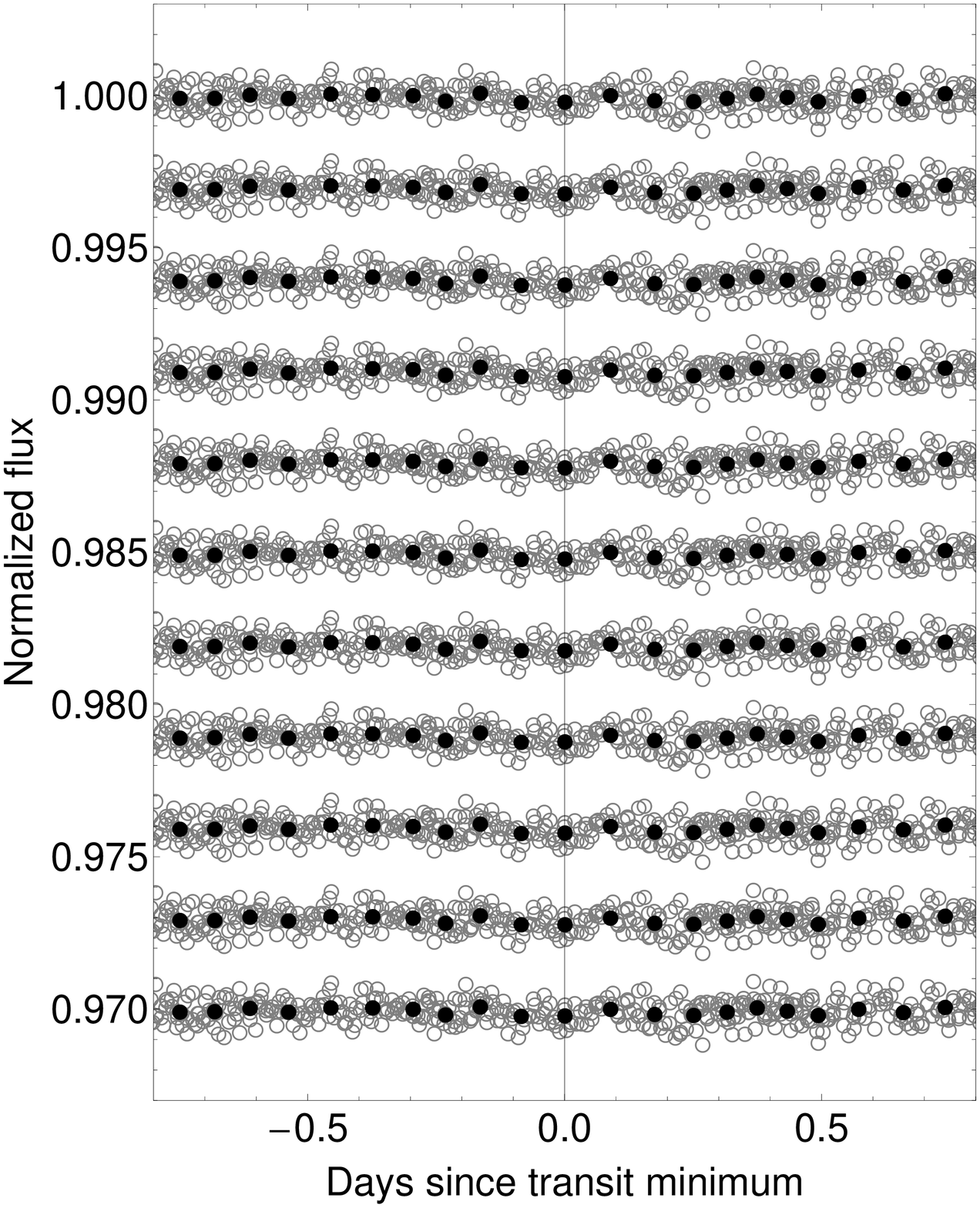}
\caption{Using the TTVs of \koib, we are able to predict the times of transit
for \koic. Assuming masses for planet b ranging from 0 (top) to 6\,$M_J$ 
(bottom) for the primary in equal steps, we show the {\it Kepler} photometry
phased upon these 11 candidate TTV ephemeres. Gray indicates the original data
and black indicates 20-point binned data. We find no evidence for
even a grazing geometry of \koic.}
\label{fig:planetc}
\end{figure}

Our analysis of the transit timing variations (TTVs) leads us to conclude that
a second planet exists in the system with an orbital period of 57\,d. Given that
the inner planet transits, and conclusion of nearly coplanar orbits from the
TTV fits, there seems a reasonable hope for detecting transits of a second
planet. However, given the fact \koic\ is Saturn-mass, it should be a gas giant
with a sizeable transit, roughly of the same size as \koib. Such a transit
would be easily spotted even by eye but there is no evidence for such events.
Fig.~\ref{fig:planetc} shows the phased data, accounting for the TTVs of
\koic\ upon the time of expected transit. Since the TTVs of \koic\ depend upon
the unknown mass of \koib, we present 11 different realizations for various
masses of planet b. In all cases no transit-like event is detectable.

Despite the fact \koic\ does not seem to transit, we initiated a search for
additional transits in the system. The light curve was searched for transits
using the Box Least-Squares method \cite{kovacs:2002}.  After removing 
the transits of \koib, we detected a significant signal (SNR$\sim14$) in the 
light curve with an apparent depth of $\sim 0.3$\,mmag, and a period of 
$P=6.7668293$\,d.  The drop in brightness had a first-to-last-contact duration, 
relative to the total period, of $q = 0.0219$, corresponding to a total duration 
of $3.6$\,hr. The transit candidate is hereby referred to as \koithree.

\paragraph*{Transit Fit of \koithree}

Initial inspection of depth, duration and period for the \koithree\ transits
suggested a physically plausible signal. Given the high significance of the 
signal, we investigated further by performing a full transit light curve fit 
accounting for the limb darkening of the star, the variable diluted light 
factors and the finite integration time of the long-cadence data, all of which 
are ignored in the BLS search.

We trim the data set to be within $\pm0.5$\,d of the expected transit times,
as computed from the BLS peak, in order to reduce the computation time. The
74 transit epochs require 74 OOT parameters to fit in conjunction with the
transit parameters themselves. This large number of parameters makes a
\multi\ fit unfeasible. Instead, we use a Markov Chain Monte Carlo (MCMC) 
routine with the Metropolis-Hastings rule. We executed two independent fits,
one using a prior on $\rho_*$ from the $\mathcal{M}_T$ fit and one with a
uniform prior on $\rho_*$ between the boundaries for a main-sequence star.

\begin{figure}[h!]
\includegraphics[width=14 cm]{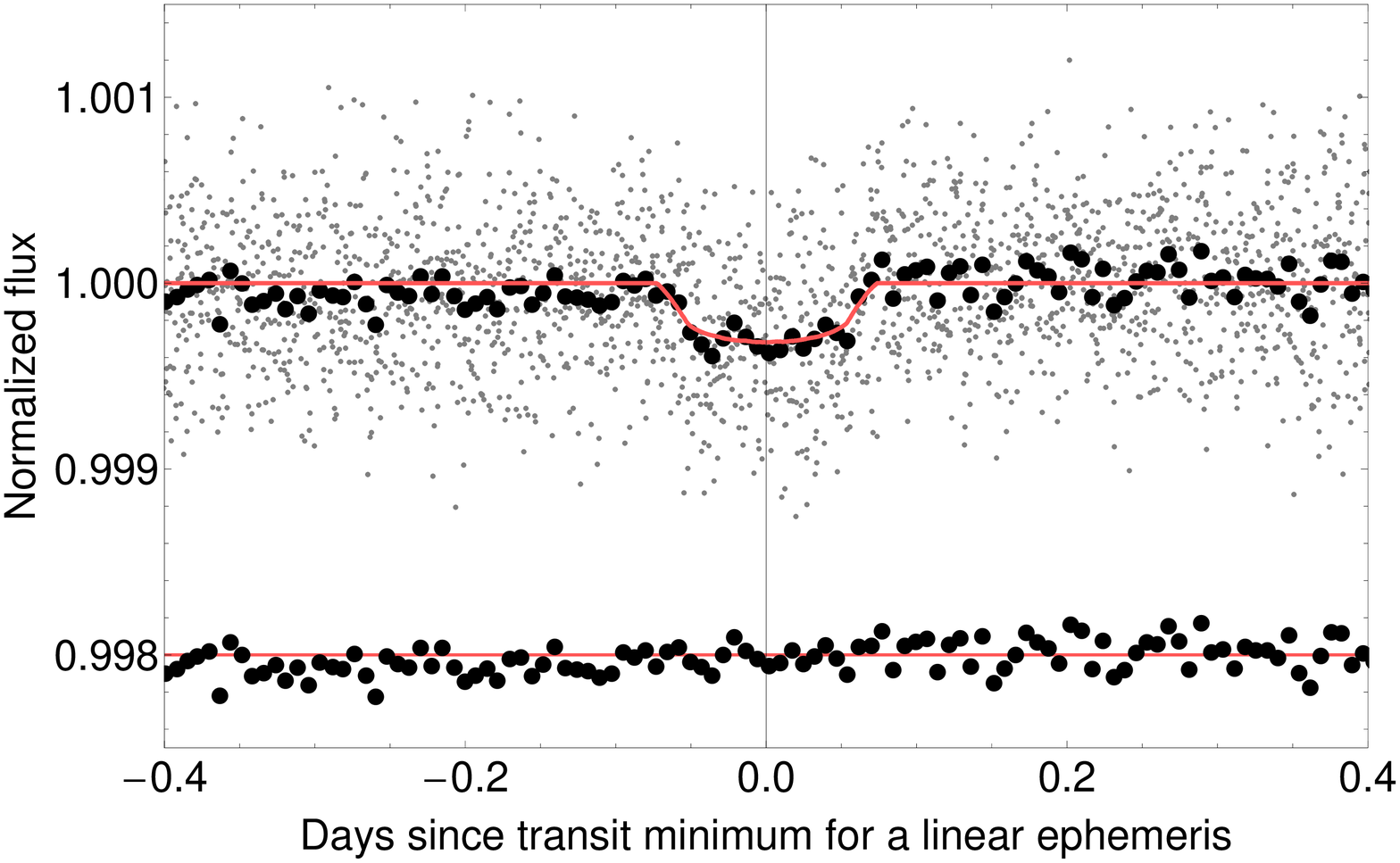}
\caption{Maximum likelihood realization of a Markov Chain Monte Carlo fit
for the transits of \koithree. Gray points show original data and black is 
20-point phase binned data. The maximum likelihood model is in red and the 
residuals are shown below offset at +0.998.}
\label{fig:planetd}
\end{figure}

With the free $\rho_*$ fit, we detect a transit corresponding to a planet of
size $(R_P/R_*)^2 = 278_{-32}^{+26}$\,ppm (8.8\,$\sigma$). With the prior
$\rho_*$ this becomes $(R_P/R_*)^2 = 274_{-28}^{+26}$\,ppm (9.9\,$\sigma$),
corresponding to a planet of size $R_P = 1.70_{-0.11}^{+0.11}$\,$R_{\oplus}$.
The maximum likelihood realization of this latter fit is shown in 
Fig.~\ref{fig:planetd} and the corresponding parameters estimates are provided
in Table~1.

As expected, the prior-$\rho_*$ fit retrieves virtually the same $\rho_*$
as derived from planet b, specifically 
$\rho_* = 1560_{-150}^{+150}$\,kg\,m$^{-3}$. The impact parameter converges to
$b = 0.39_{-0.12}^{+0.19}$ corresponding to $i=(88.55_{-0.49}^{+0.69})^{\circ}$.
Curiously though, releasing this prior yields
$\rho_* = 1820_{-490}^{+660}$\,kg\,m$^{-3}$, $b=0.02_{-0.44}^{+0.44}$ and 
$i=(88.92_{-1.26}^{+0.75})^{\circ}$. In other words, without any prior
information, the light curve yields a consistent stellar density (within 
0.6\,$\sigma$), highly indicative that \koithree\ orbits the same star as 
\koib\ with negligible eccentricity.

Due to the low signal-to-noise, we were not able to determine individual transit 
times or durations for this object. We note that the expected TTVs of \koib\ due 
to \koithree\ is less than 1\,second in amplitude and thus undetectable with the 
current data. Further, the expected TTVs of \koithree\ are around 10\,s, which 
are again too small to detect. Without TTVs, we cannot causally link the object 
to be transiting the same host star as \koithree\ (it could be transiting a 
background star). Even assuming it was in the same system stability limits allow 
the object to be as massive as 100\,$M_J$ and still be stable. Although formally 
unconfirmable, the derived $\rho_*$ suggests it is likely associated with the 
same star.

\paragraph*{A Search for an Exomoon}

As discussed earlier, the observed TTVs cannot be adequately explained by
an exomoon and only a second planet in the 5:3 resonance offers a valid
solution. The presence of this second planet complicates our search for an 
exomoon. Further, the presence of stellar activity makes spot crossings and 
transit distortions probable, further exacerbating our search for a moon. 
Despite this, we here describe our efforts to search for an extrasolar moon.

{\it Fitting for a Moon Eclipse After Removing the TTVs, $\mathcal{M}_{MT1,M0}$.}
The first attempt we made was to forcibly remove the best-fit TTVs from the
time series and then fit for a zero-mass moon. This fit allows for a finite
radius moon and thus is merely a search for the moon eclipse. We perform
two versions of the model fit; one setting the exomoon radius to zero
and one allowing the parameters $R_S/R_P$, $(\rho_P)^{2/3}$, $P_S$, $\phi_S$,
$i_S$ and $\Omega_S$ to be freely varied (see \cite{luna:2011} for
various definitions of these terms). The two models are performed so that
we have a null model to compare against.

The results yield $\log\mathcal{Z} = (12001.34 \pm 0.37)$ for the null fit
and $\log\mathcal{Z} = (12025.89 \pm 0.24)$ for the moon-transit fit, or a
6.7-$\sigma$ preference for the moon-transit model. Whilst certainly above
our statistical significance threshold, one should recall that the star is
active and these fitted events could merely be star spot crossings or
activity-related events. Such events would be poorly sampled with the 30\,minute
cadence of the current observations though.

The mass-ratio of the planet and the star may be determined using the light
curve alone, as described in \cite{weighing:2010}. For our estimate of 
the stellar mass, this allows us to compute $M_P$ directly. The results find 
that $M_P > 20$\,$M_J$ for all modes, which exceeds the 6\,$M_J$ stability
limit imposed on the system. We therefore conclude that none of these modes
are genuine and most likely due to the presence of stellar activity. The likely
presence of these spots also prevent us placing an upper excluded limit on a 
putative exomoon radius.

\begin{figure*}
\begin{center}
\includegraphics[width=12 cm]{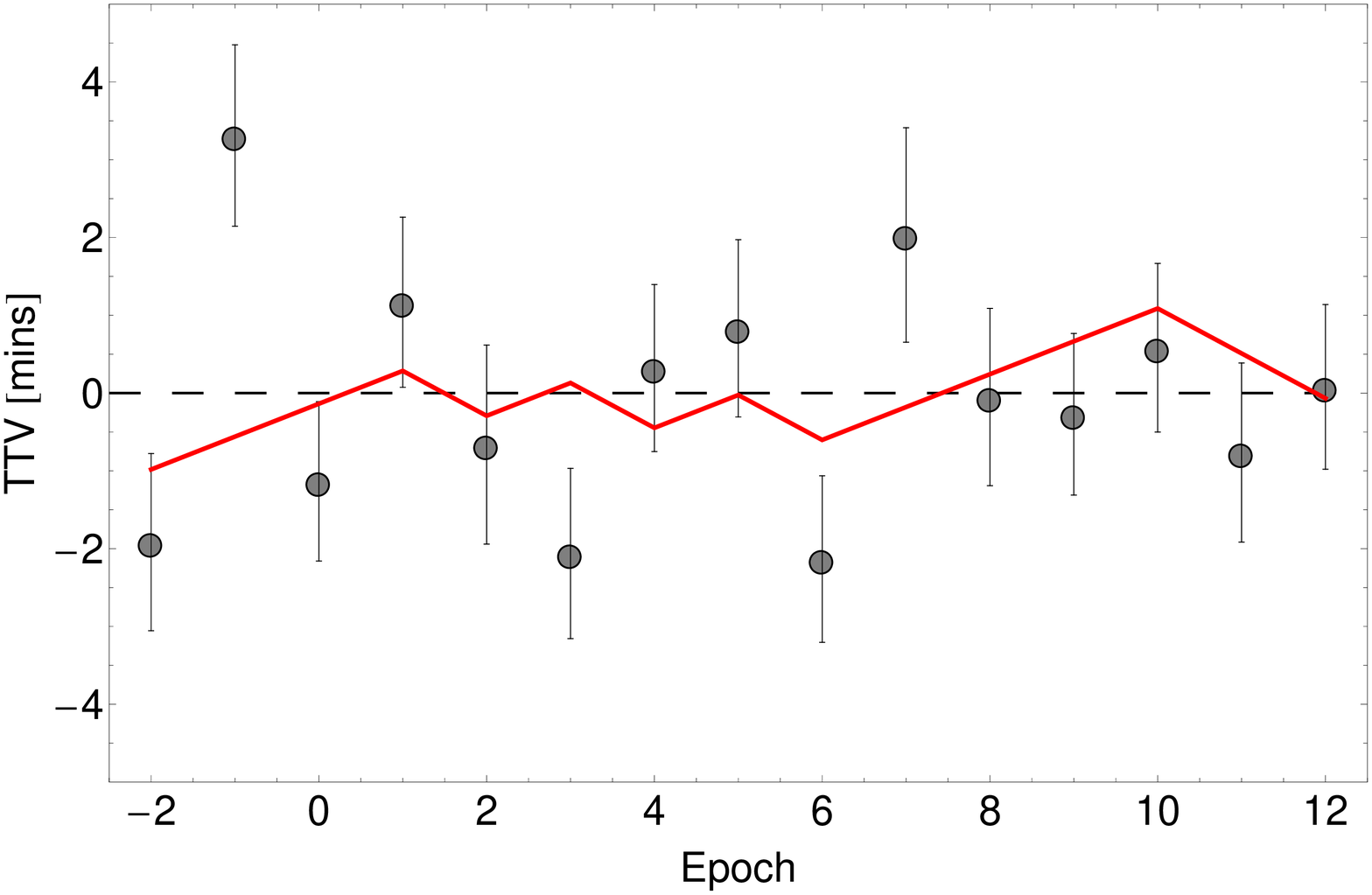}
\caption{Maximum a-posteriori moon fit to the TTVs of \koib\ (using model
$\mathcal{M}_{MT2,R0}$), after removing the maximum likelihood TTVs due to the
second planet, \koic. We find no significant improvement by including a moon.
}
\label{fig:TTVresiduals}
\end{center}
\end{figure*}

\begin{figure*}
\begin{center}
\includegraphics[width=12 cm]{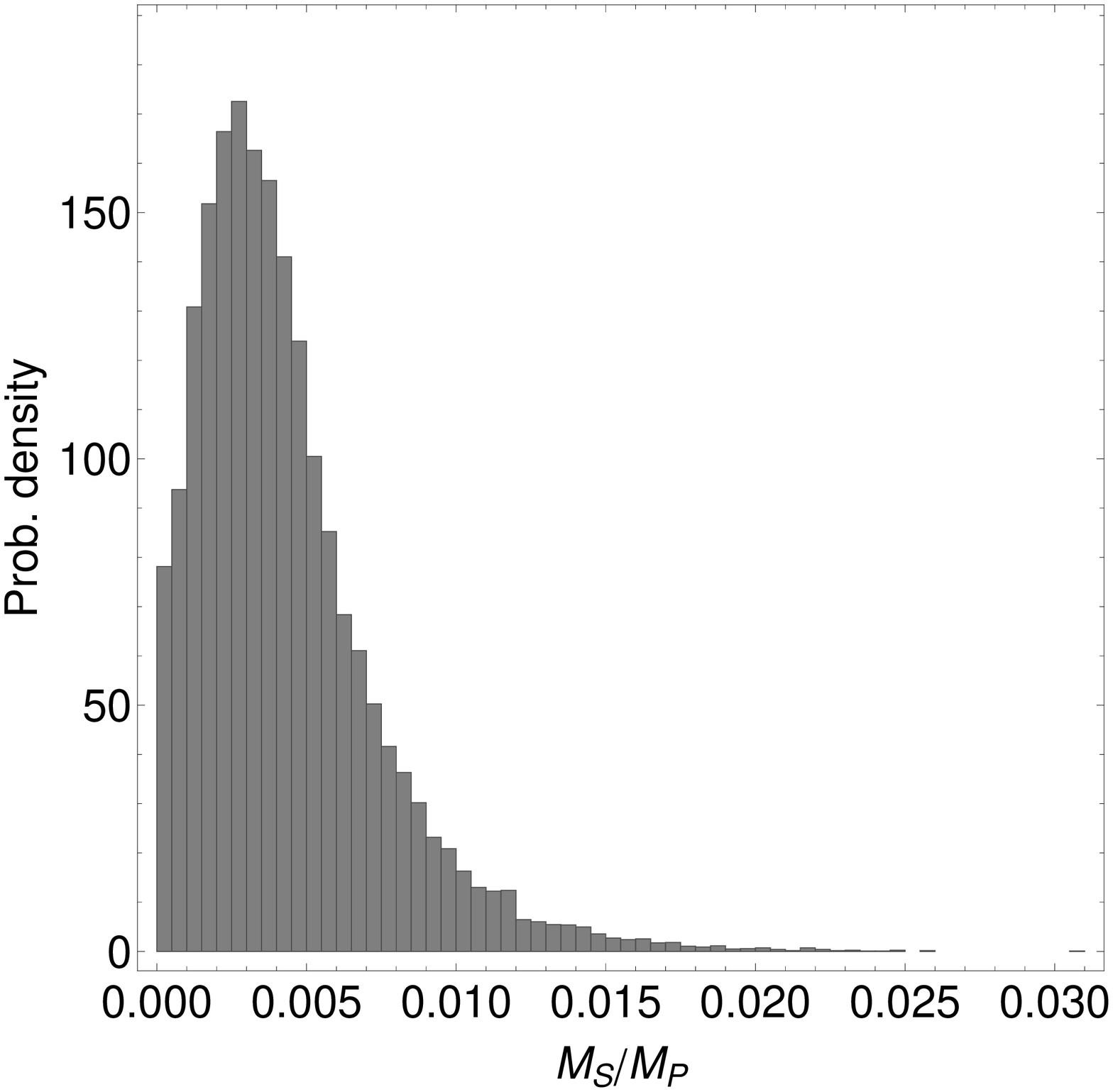}
\caption{Posterior distribution for the mass of an exomoon relative to
the mass of \koib\, as computed using model $\mathcal{M}_{MT2,R0}$, marginalized 
over the entire prior volume. We estimate a 3-$\sigma$ upper limit of
$M_S/M_P < 0.021$ for this planet.
}
\label{fig:moonlimits}
\end{center}
\end{figure*}

{\it Fitting for a Moon After Removing Planet TTVs, $\mathcal{M}_{MT2,R0}$.}
We also tried removing the maximum likelihood TTVs from the planet-fit of
\koic. This is therefore a fit on the residual TTVs for a moon signal. Since
the eclipse signal of the moon is likely unreliable due to stellar activity, we
limit this search to a model, where the moon is a point-mass (i.e. $R_S=0$);
model $\mathcal{M}_{MT2,R0}$. We found this model unable to locate a 
significantly improved fit with $\Delta\log\mathcal{Z} = -(1.55 \pm 0.44)$, 
relative to the null hypothesis. To illustrate this, Fig.~\ref{fig:TTVresiduals} 
shows the TTV residuals after removing the TTVs of \koic\ along with the maximum 
likelihood model TTVs from $\mathcal{M}_{MT2,R0}$. Although no exomoon is 
detected, we can use the results to place upper limits on a putative exomoon 
mass. This is particularly valuable given that radius limits are not possible 
due to the likely presence of spots. Fig.~\ref{fig:moonlimits} provides the 
corresponding mass limits, excluding $M_S/M_P < 0.021$ to 3-$\sigma$ confidence.

\paragraph*{TTV Constraints on Additional Planets}
When the computed TTVs corresponding to s1 are subtracted from the measured 
TTVs, this leaves a small residual signal with a $\simeq$1 minute amplitude. 
This is comparable to the measurement errors.

The small amplitude of the TTV residuals can be used to place limits on the 
presence of additional planets in the system. Given that \koib\ and \koic\ have 
nearly coplanar orbits, we tested a case in which the additional planet was 
placed in the invariant plane of the two confirmed planets. The orbits were 
followed by an $N$-body integrator to see whether the computed TTVs are 
consistent with the residuals. 

The results are illustrated in Figs. \ref{surv} and \ref{surv2}. The small 
amplitude of residual TTVs provides an useful constraint on the third planet's 
mass and orbit. They rule out, for example, a Jupiter-mass planet on low-$e$ 
orbit with $0.05<a<0.5$~AU. 

Additional constraints can be obtained from the stability requirements. For 
example, a low-mass planet with $e \simeq 0$ and $i \simeq 0$ should have 
$|a-a_c|/a_c > C (M_c/M_*)^{2/7}$, where $M_c$ and $a_c$ are the mass and 
semimajor axis of \koic, for the system to be stable \cite{wisdom:1980}, where 
$C\simeq1.5$ (e.g., \cite{quillen:2006,chiang:2009}). For 
$M_c/M_*=4\times10^{-4}$, this gives $|a-a_c|/a_c > 0.160$. 

A more accurate stability criterion was derived in \cite{mustill:2012}. For 
$M_c/M_* = 4\times 10^{-4}$ and $e_c = 0.015$, this criterion gives 
$|a-a_c|/a_c > 1.8 e_c^{1/5} (M_c/M_*)^{1/5} = 0.162$. The difference between 
the two criteria is therefore negligible in our case. As the mean motion 
resonances become wider with planet's eccentricity, the eccentricity-dependent 
criterion of \cite{mustill:2012} should be used for larger $e$.



\clearpage

\begin{figure}
\centering
\includegraphics[width=14.cm]{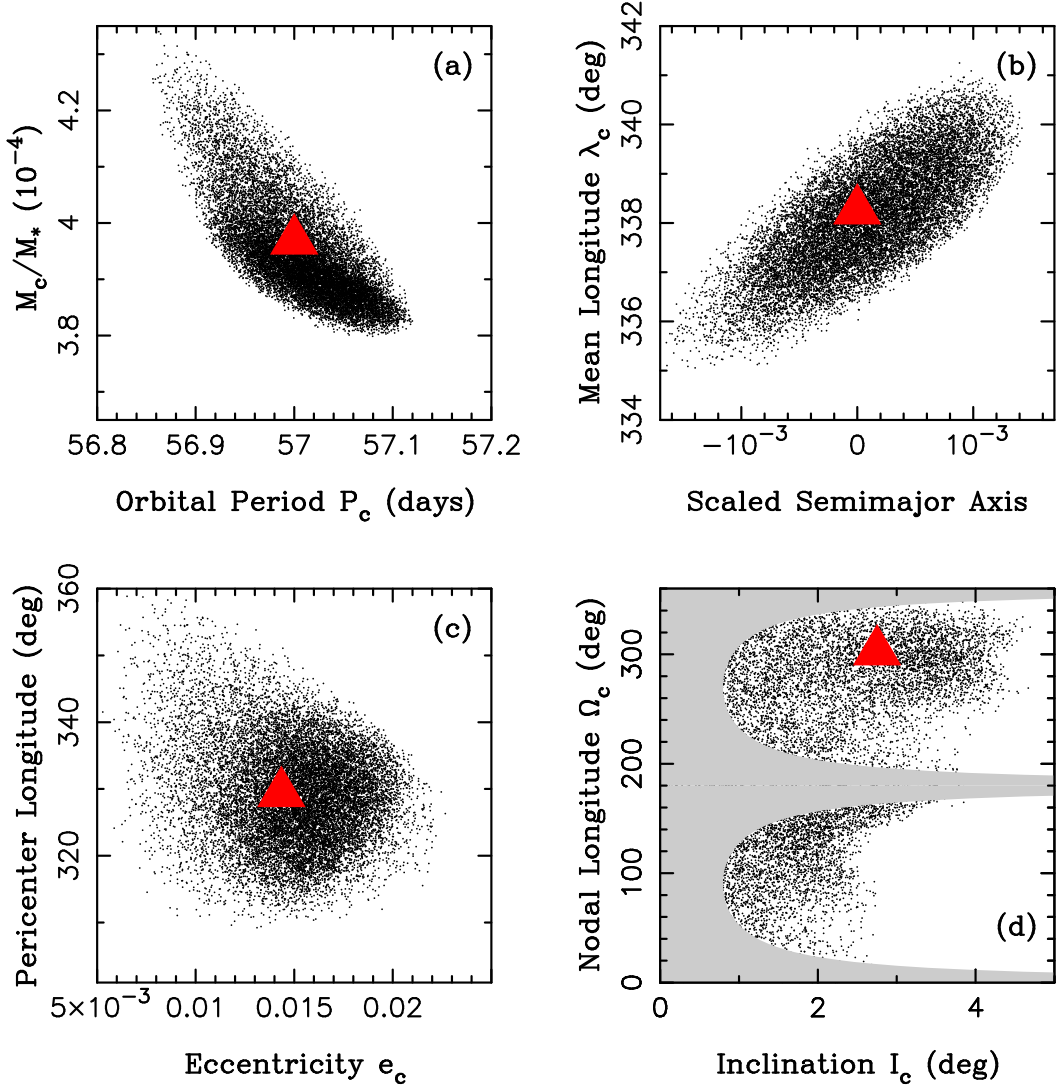}
\caption{Confidence intervals. The dots denote the 99\% confidence area around 
s1 (red triangles). We sampled the general neighborhood of s1, determined 
$\chi^2$ for each parameter set (3~million parameter sets in total) and plotted 
a dot if $\chi^2<\chi^2_{\rm min}+\Delta\chi^2(99\%)=23.5$. All parameters shown 
here are well constrained, including the orbital inclination of \koic\ (the gray 
area in (d) is ruled out by the lack of \koic's transits). This result was 
obtained while fixing $e_b=0$. A nearly identical result was obtained by letting 
$e_b$ (and $\varpi_b$) vary. In that case, $\chi^2<\chi^2_{min}+
\Delta\chi^2(99\%)$ with $\Delta\chi^2(99\%)=16.9$ for $15-9=6$ DOF gives 
$e_b<0.02$, leaving $\varpi_b$ unconstrained. The scaled semimajor axis in panel 
(b) is defined as $(a-a_c)/a_c$, where $a_c$ is the best-fit semimajor axis 
value for solution s1.
}
\label{fig2}
\end{figure}

\begin{figure}[h!]
\includegraphics[width=14.cm]{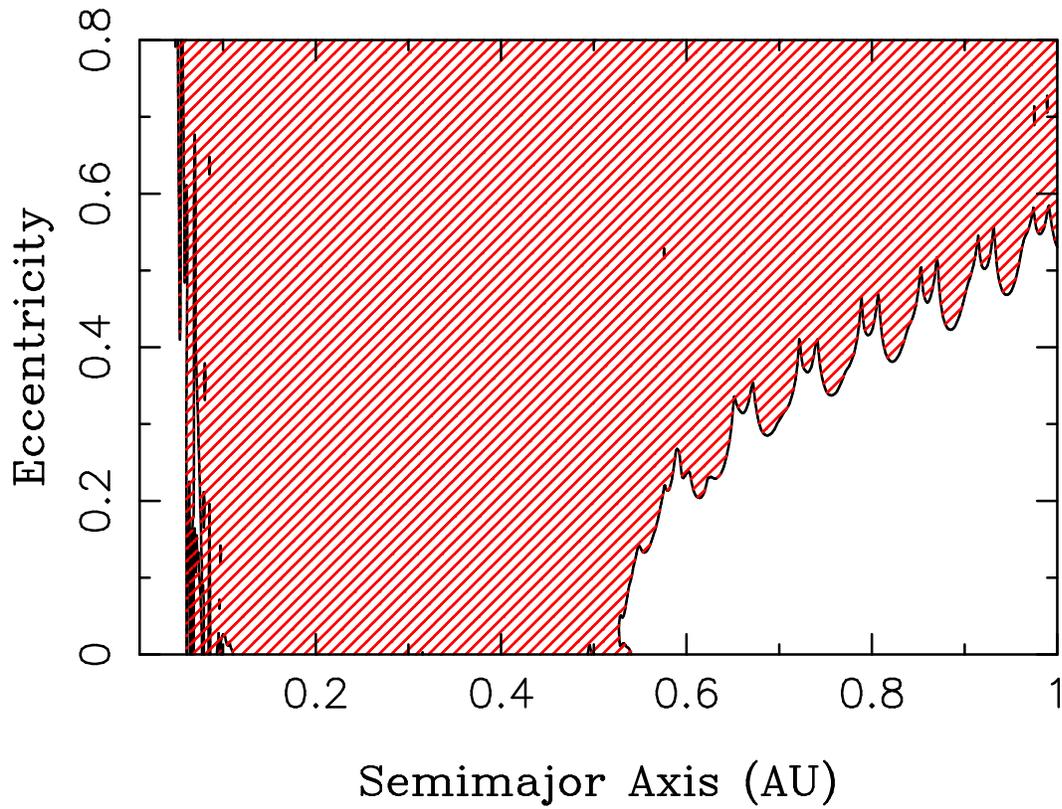}
\caption{Constraints on the presence of an additional Jupiter-mass planet. The 
dashed area shows where the Jupiter-mass planet would induce the TTV amplitude 
of \koib\ in excess of 1 minute.  If the residual TTVs are caused by a 
Jupiter-mass planet, this planet should have $a$ and $e$ near the boundary of 
the shaded area.}
\label{surv}
\end{figure}

\begin{figure}[h!]
\includegraphics[width=14.cm]{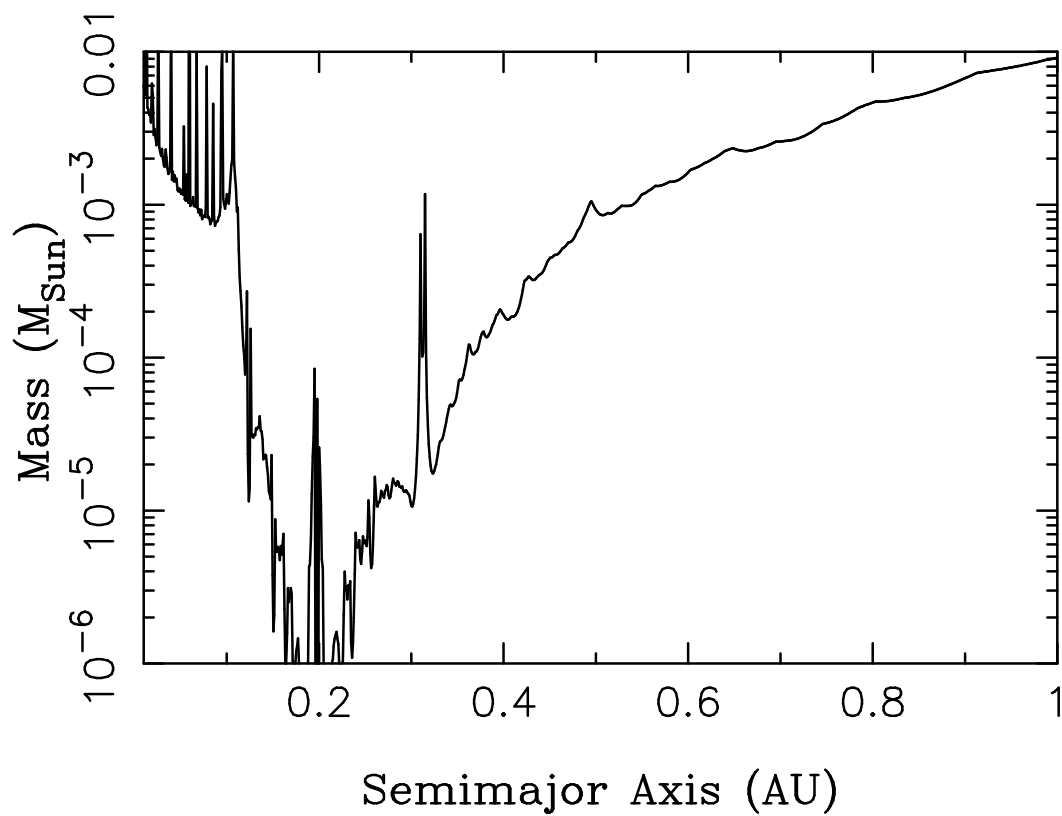}
\caption{Constraints on the presence of an additional planet on circular orbit. 
The line shows where the planet induces the TTV amplitude of 1 minute. Note 
that the low-mass planet on a circular orbit exterior to \koib\ cannot explain 
the TTV residuals, because such a planet would be perturbed by \koic\ and would 
be dynamically unstable.}
\label{surv2}
\end{figure}

\clearpage


\begin{table}
\centering
{
\begin{tabular}{llll}
\hline\hline 
Parameter & $\mathcal{M}_P$ & $\mathcal{M}_V$ \& $\mathcal{M}_T$ & $\mathcal{M}_{M*}$ \\ [0.5ex] 
\hline
$p$\,$^{\ddagger}$ & $\mathcal{U}[0,0.25]$ & $\mathcal{U}[0,0.25]$ & $\mathcal{U}[0,0.25]$ \\
$[\rho_{*}^{\mathrm{circ}}]^{2/3}$\,[kg$^{2/3}$\,m$^{-2}$]$^{\ddagger}$ & $\mathcal{U}[7.6499,6097.85]$ & $\mathcal{U}[7.6499,6097.85]$ & $\mathcal{U}[7.6499,6097.85]$ \\
$b_P$\,$^{\ddagger}$ & $\mathcal{U}[0,1.25]$ & $\mathcal{U}[0,1.25]$ & $\mathcal{U}[0,1.25]$ \\
$P_P$\,[days] & $\mathcal{U}[P_{P}^{*}-1,P_{P}^{*}+1]$ & $\mathcal{N}[P_{P}^{*},0.0002]$ & $\mathcal{U}[P_{P}^{*}-1,P_{P}^{*}+1]$ \\
$\tau_0$ & $\mathcal{U}[\tau_{0}^{*}-1,\tau_{0}^{*}+1]$ & N/A & $\mathcal{U}[\tau_{0}^{*}-1,\tau_{0}^{*}+1]$ \\
$a_1$ & $\mathcal{U}[150,250]$ & $\mathcal{U}[150,250]$ & $\mathcal{U}[150,250]$ \\ 
$a_2$ & $\mathcal{U}[0.8,1.3]$ & $\mathcal{U}[0.8,1.3]$ & $\mathcal{U}[0.8,1.3]$ \\
$OOT_n$ & $\mathcal{U}[0.95,1.05]$ & $\mathcal{U}[0.95,1.05]$ & $\mathcal{U}[0.95,1.05]$ \\
$OOT_{-1}$ & $\mathcal{U}[8589.6487,9493.8223]$ & $\mathcal{U}[8589.6487,9493.8223]$ & $\mathcal{U}[8589.6487,9493.8223]$ \\
$\tau_n$ & N/A & $\mathcal{U}[\tau_{n}^{*}-1,\tau_{n}^{*}+1]$ & N/A \\
$e_P$ & $\delta[0]$ & $\delta[0]$ & $\delta[0]$ \\
$\omega_P$\,[rads] & N/A & N/A & N/A \\
$u_1$ & $\delta[0.3542]$ & $\delta[0.3542]$ & $\delta[0.3542]$ \\
$(u_1 + u_2$) & $\delta[0.7140]$ & $\delta[0.7140]$ & $\delta[0.7140]$ \\
$\sigma_W$ & $\mathcal{J}'[\sigma_{\mathrm{photo}},10 \sigma_{\mathrm{photo}}]$ & $\mathcal{J}'[\sigma_{\mathrm{photo}},10 \sigma_{\mathrm{photo}}]$ & $\mathcal{J}'[\sigma_{\mathrm{photo}},10 \sigma_{\mathrm{photo}}]$ \\
\hline
$(R_S/R_P)$ & N/A & N/A & $\mathcal{U}[0,1]$ \\
$P_S$\,[days] & N/A & N/A & $\mathcal{U}[0.083,19.3944]$ \\
$\phi_S$\,[rads] & N/A & N/A & $\mathcal{U}[0,2\pi]$ \\
$[\rho_P]^{2/3}$\,[kg$^{2/3}$\,m$^{-2}$] & N/A & N/A & $\mathcal{U}[18.4982,920.9760]$ \\
$i_{S}$\,[rads] & N/A & N/A & $\mathcal{U}[0,2\pi]$ \\
$\Omega_{S}$\,[rads] & N/A & N/A & $\mathcal{U}[-\pi/2,\pi/2]$ \\
$[\rho_{S}]^{2/3}$\,[kg$^{2/3}$\,m$^{-2}$]$^{\dagger\times}$ & N/A & N/A & $\mathcal{U}[0,920.9760]$ \\
$e_{S}$ & $\delta[0]$ & $\delta[0]$ & $\delta[0]$ \\
$\omega_{S}$\,[rads] & N/A & N/A & N/A \\ [1ex]
\hline\hline 
\end{tabular}
	}
\caption{Priors used for various free parameters in the \multi\ fits.
$\mathcal{U}[x,y]$ is a uniform prior between $x$ and $y$.
$\mathcal{N}[x,y]$ is a Gaussian prior with mean $x$ and standard deviation $y$.
$\delta[x]$ is a delta-function prior centered on $x$. 
$\mathcal{J}'[x,y]$ is a modified Jeffrey's prior with an inflection point at 
$x$ and a maximum limit of $y$.
We use the replacements $\tau_{0}^{*} = 2455053.2815$\,BJD$_{\mathrm{UTC}}$, 
$P_{P}^{*} = 33.6013$\,d and $\tau_{n}^{*} = \tau_{0}^{*} + n P_{P}^{*}$.
$\sigma_{\mathrm{photo}}$ is the median photometric error outputted from
the \emph{Kepler} pipeline for the time series under analysis.
$^{\ddagger}$ = for $\mathcal{M}_V$ fits these terms are independently fitted to 
each transit epoch.
N/A = parameter is fixed to some arbitrary value, since it has no 
influence on the fits. 
$^{\dagger}$ = for $\mathcal{M}_{M,R0}$ and $\mathcal{M}_{MT2,R0}$, this term is 
replaced with $(M_S/M_P)$ and treated as uniform between 0 and 1.
$^{\times}$ = for $\mathcal{M}_{MT1,M0}$, this term is fixed to zero to 
represent a zero-mass moon.}
\label{tab:priors} 
\end{table}

\begin{table}
\centering
{
\begin{tabular}{ll}
\hline\hline 
Model & $\log\mathcal{Z}$ \\ [0.5ex] 
\hline
\emph{Original data} \\
\hline
$\mathcal{M}_P$ & $(10981.35 \pm 0.35)$ \\
$\mathcal{M}_T$ & $\geq(11948.82 \pm 0.28)$ \\
$\mathcal{M}_V$ & $(11782.73 \pm 0.61)$ \\
$\mathcal{M}_M$ & $(11313.56 \pm 0.22)$ \\
$\mathcal{M}_{M,R0}$ & $(11403.03 \pm 0.22)$ \\ 
\hline
\emph{All TTVs removed} \\
\hline
$\mathcal{M}_{MT1,\mathrm{null}}$ & $(12001.34 \pm 0.37)$ \\
$\mathcal{M}_{MT1,M0}$ & $(12025.89 \pm 0.24)$ \\
\hline
\emph{Planet TTVs removed} \\
\hline
$\mathcal{M}_{MT2,\mathrm{null}}$ & $(12006.43 \pm 0.37)$ \\ 
$\mathcal{M}_{MT2,R0}$ & $(12004.90 \pm 0.23)$ \\ [1ex]
\hline\hline 
\end{tabular}
	}
\caption{Bayesian evidences for the planetary system family of models 
fitted to the \koi\ photometry. The data favor model $\mathcal{M}_T$, a
planet-only model with transit timing variations (TTV).}
\label{tab:evidences}
\end{table}

\begin{table}
\centering
{
\begin{tabular}{llll}
\hline\hline 
Epoch & $\mathrm{OOT}_{i}$ & $\tau_i$\,[BJD$_{\mathrm{UTC}}$] & TTV\,[mins] \\ [0.5ex] 
\hline
-2 & $0.999732_{-0.000020}^{+0.000020}$ & $2454986.09325_{-0.00079}^{+0.00079}$ & $20.8_{-1.1}^{+1.1}$  \\
-1 & $9043.45_{-0.31}^{+0.30}$          & $2455019.69845_{-0.00081}^{+0.00081}$ & $26.4_{-1.2}^{+1.2}$  \\
0  & $1.000150_{-0.000020}^{+0.000020}$ & $2455053.29450_{-0.00071}^{+0.00073}$ & $18.7_{-1.0}^{+1.1}$  \\
1  & $0.995315_{-0.000020}^{+0.000020}$ & $2455086.86287_{-0.00076}^{+0.00075}$ & $-28.8_{-1.1}^{+1.1}$ \\
2  & $0.999953_{-0.000022}^{+0.000021}$ & $2455120.43870_{-0.00089}^{+0.00087}$ & $-65.5_{-1.3}^{+1.3}$ \\
3  & $1.004089_{-0.000041}^{+0.000040}$ & $2455154.07497_{-0.00076}^{+0.00076}$ & $-15.2_{-1.1}^{+1.1}$ \\
4  & $0.999651_{-0.000021}^{+0.000020}$ & $2455187.68561_{-0.00075}^{+0.00076}$ & $-1.9_{-1.1}^{+1.1}$  \\
5  & $0.999807_{-0.000021}^{+0.000020}$ & $2455221.33536_{-0.00079}^{+0.00081}$ & $67.8_{-1.1}^{+1.2}$  \\
6  & $1.000474_{-0.000020}^{+0.000021}$ & $2455254.90533_{-0.00074}^{+0.00078}$ & $22.7_{-1.1}^{+1.1}$  \\
7  & $0.999937_{-0.000020}^{+0.000020}$ & $2455288.46626_{-0.00096}^{+0.00096}$ & $-35.6_{-1.4}^{+1.4}$ \\
8  & $0.999864_{-0.000020}^{+0.000020}$ & $2455322.05975_{-0.00079}^{+0.00081}$ & $-46.9_{-1.1}^{+1.2}$ \\
9  & $0.999749_{-0.000020}^{+0.000020}$ & $2455355.66579_{-0.00072}^{+0.00072}$ & $-40.1_{-1.0}^{+1.0}$ \\
10 & $0.999834_{-0.000021}^{+0.000020}$ & $2455389.34350_{-0.00075}^{+0.00076}$ & $69.8_{-1.1}^{+1.1}$  \\
11 & $0.999762_{-0.000020}^{+0.000020}$ & $2455422.92851_{-0.00080}^{+0.00079}$ & $46.3_{-1.2}^{+1.1}$  \\
12 & $1.000111_{-0.000020}^{+0.000020}$ & $2455456.48567_{-0.00074}^{+0.00076}$ & $-17.3_{-1.1}^{+1.1}$  \\ [1ex]
\hline\hline 
\end{tabular}
	}
\caption{Epoch-specific fitted parameters for the \koi\ system. Results 
computed from the weighted posteriors resulting from model $\mathcal{M}_T$, 
using \multi. TTVs relative to maximum a-posteriori linear ephemeris derived in
model fit $\mathcal{M}_P$; $P_P = 33.6013506$\,d and $\tau_0 = 2455053.2815010$.
Physical system parameters are provided in Table~1.}
\label{tab:TTVs}
\end{table}

\begin{table}
\centering
{
\begin{tabular}{llllll}
\hline\hline 
Epoch & $\mathrm{OOT}_{i}$ & $T_{14}$\,[mins] & TDV [mins] & $\tau_i$\,[BJD$_{\mathrm{UTC}}$] & TTV\,[mins] \\ [0.5ex] 
\hline
-2 & $0.999716_{-0.000043}^{+0.000043}$ & $249_{-20}^{+18}$      & $-2_{-20}^{+18}$       & $2454986.0935_{-0.0017}^{+0.0018}$ & $21.2_{-2.4}^{+2.5}$ \\
-1 & $9043.31_{-0.55}^{+0.57}$          & $239.3_{-8.0}^{+17.5}$ & $-11.3_{-8.0}^{+17.5}$ & $2455019.6976_{-0.0015}^{+0.0016}$ & $25.1_{-2.2}^{+2.4}$ \\
0  & $1.000152_{-0.000037}^{+0.000038}$ & $242.3_{-5.4}^{+7.0}$  & $-8.3_{-5.4}^{+7.0}$   & $2455053.2943_{-0.0013}^{+0.0013}$ & $18.4_{-1.9}^{+1.9}$ \\
1  & $0.995332_{-0.000040}^{+0.000039}$ & $255_{-12}^{+19}$      & $5_{-12}^{+19}$        & $2455086.8629_{-0.0015}^{+0.0015}$ & $-28.7_{-2.1}^{+2.2}$ \\
2  & $0.999948_{-0.000033}^{+0.000035}$ & $256.5_{-7.8}^{+18.4}$ & $6.0_{-7.8}^{+18.4}$   & $2455120.4391_{-0.0014}^{+0.0013}$ & $-64.9_{-2.0}^{+1.9}$ \\
3  & $1.00403_{-0.00011}^{+0.00011}$    & $246.5_{-8.4}^{+10.8}$ & $-4.0_{-8.4}^{+10.8}$  & $2455154.0749_{-0.0020}^{+0.0019}$ & $-15.3_{-2.9}^{+2.8}$ \\
4  & $0.999660_{-0.000038}^{+0.000037}$ & $249.3_{-6.3}^{+10.5}$ & $-1.3_{-6.3}^{+10.5}$  & $2455187.6856_{-0.0014}^{+0.0014}$ & $-1.9_{-2.0}^{+2.0}$ \\
5  & $0.999811_{-0.000035}^{+0.000034}$ & $251.7_{-5.9}^{+9.8}$  & $1.2_{-5.9}^{+9.8}$    & $2455221.3355_{-0.0012}^{+0.0012}$ & $68.0_{-1.8}^{+1.8}$ \\
6  & $1.000453_{-0.000033}^{+0.000033}$ & $244.3_{-5.5}^{+8.9}$  & $-6.2_{-5.5}^{+8.9}$   & $2455254.9051_{-0.0012}^{+0.0012}$ & $22.4_{-1.7}^{+1.7}$ \\
7  & $0.999924_{-0.000043}^{+0.000043}$ & $270_{-33}^{+22}$      & $20_{-33}^{+22}$       & $2455288.4673_{-0.0019}^{+0.0020}$ & $-34.0_{-2.7}^{+2.9}$ \\
8  & $0.999868_{-0.000037}^{+0.000035}$ & $254.2_{-5.4}^{+9.9}$  & $3.7_{-5.4}^{+9.9}$    & $2455322.0599_{-0.0012}^{+0.0012}$ & $-46.7_{-1.8}^{+1.7}$ \\
9  & $0.999741_{-0.000041}^{+0.000041}$ & $247.6_{-6.8}^{+8.9}$  & $-2.9_{-6.8}^{+8.9}$   & $2455355.6657_{-0.0015}^{+0.0015}$ & $-40.3_{-2.2}^{+2.2}$ \\
10 & $0.999855_{-0.000039}^{+0.000038}$ & $242.3_{-7.4}^{+12.4}$ & $-8.3_{-7.4}^{+12.4}$  & $2455389.3435_{-0.0014}^{+0.0014}$ & $69.8_{-2.0}^{+2.0}$ \\
11 & $0.999786_{-0.000035}^{+0.000036}$ & $267_{-20}^{+17}$      & $16_{-20}^{+17}$       & $2455422.9284_{-0.0014}^{+0.0014}$ & $46.2_{-2.0}^{+2.0}$ \\
12 & $1.000123_{-0.000040}^{+0.000040}$ & $251.2_{-8.0}^{+13.5}$ & $0.7_{-8.0}^{+13.5}$   & $2455456.4856_{-0.0014}^{+0.0015}$ & $-17.4_{-2.0}^{+2.2}$ \\ [1ex]
\hline\hline 
\end{tabular}
	}
\caption{Epoch-specific fitted parameters for the \koi\ system. Results 
computed from the weighted posteriors resulting from model $\mathcal{M}_V$, 
using \multi. TDVs relative to median duration of all duration realizations
in $\mathcal{M}_V$; $T_{14} = 250.5$\,mins.
Physical system parameters are provided in Table~1.}
\label{tab:TDVs}
\end{table}

\begin{table}
\centering
{
\begin{tabular}{llll}
\hline\hline 
Hypothesis & $\log\mathcal{Z}$ & $\sigma$ Preference over $\mathcal{H}_{P}$ & Blend analysis conclusion \\ [0.5ex] 
\hline
$\mathcal{H}_{P}$ & $(12000.13 \pm 0.37)$ & - & Plausible \\
$\mathcal{H}_{EB,33.6}^{c}$ & $(12000.95 \pm 0.37)$ & $(0.8 \pm 0.4)$ & Plausible \\
$\mathcal{H}_{EB,33.6}^{e}$ & $(12004.29 \pm 0.35)$ & $+(2.4 \pm 0.2)$ & Plausible \\ 
$\mathcal{H}_{EB,67.2}^{c}$ & $(11977.45 \pm 0.39)$ & $-(6.41 \pm 0.08)$ & Excluded \\
$\mathcal{H}_{EB,67.2}^{e}$ & $(11867.05 \pm 0.36)$ & $-(16.13 \pm 0.03)$ & Excluded \\ [1ex]
\hline\hline 
\end{tabular}
	}
\caption{Bayesian evidences for a blend family of hypotheses
fitted to the \koi\ photometry. Out of the various blend scenarios tried, we
find the light curve shape has no significant preference between hypotheses
$\mathcal{H}_{P}$, $\mathcal{H}_{EB,33.6}^{c}$ and $\mathcal{H}_{EB,33.6}^{e}$.}
\label{tab:vetting}
\end{table}


\begin{thebibliography}{99}
\bibitem{tdv} 
To demonstrate this we generated $10^6$ realizations in the neighborhood of s2 
and selected those consistent with the observed TTVs (as defined by 
$\chi^2<\chi_{\rm min,s2}^2+\delta \chi^2(99\%)$). 
We found that the mass and orbit were well constrained from 
TTVs with $M_c \simeq 1.8\times10^{-3}M_*$, $81.5<P_c<82.3$\,d, 
$0.02<e_c<0.05$ and $7^\circ<I_c<12^\circ$. We then calculated TDVs for the 
selected realizations and found that all were inconsistent with the 
measurements by $>3$\,$\sigma$. In contrast, the realizations near s2 that were 
consistent with the observed TDVs had $I_c<5^\circ$, and were inconsistent with 
the observed TTVs.
\bibitem{three} We attempted to fit the observed TTVs with a three-planet 
system. The obtained solutions recovered the two-planet case corresponding to s1 
or s2, with the third planet being either too small or too distant to produce 
significant TTVs.
\end{thebibliography}
\end{document}